\documentclass[11pt]{article}
\pdfoutput=1
\usepackage{amsmath}
\usepackage{amssymb}
\usepackage{graphicx,bbm,mathrsfs}
\usepackage{nicefrac}
\usepackage{bbm}
\usepackage{geometry}       
\usepackage{changepage}
\usepackage{slashed}

\usepackage{jheppub}     

\usepackage{dcolumn}   
\usepackage{bm}        
\usepackage{graphicx,mathrsfs}
\usepackage{nicefrac}
\usepackage{multirow}
\usepackage{color}

\def\ie{{\it i.e.}}

\newcommand{\be}{\begin{equation}}  
\newcommand{\ee}{\end{equation}}  
\newcommand{\bea}{\begin{eqnarray}}  
\newcommand{\eea}{\end{eqnarray}}  

\renewcommand{\O}{\mathcal O}

\begin{document}

\vspace*{1.2cm}

\begin{center}

\thispagestyle{empty}
{\Large\bf 
Shining Light on Polarizable Dark Particles
}\\[10mm]

\renewcommand{\thefootnote}{\fnsymbol{footnote}}

{\large  Sylvain~Fichet  
\footnote{sylvain@ift.unesp.br} }\\[10mm]

\addtocounter{footnote}{-1}

{\it  ICTP South American Institute for Fundamental Research, Instituto de Fisica Teorica,\\
Sao Paulo State University, Brazil \\
}

\vspace*{12mm}

{  \bf  Abstract }
\end{center}

We investigate the possibilities of searching for a self-conjugate polarizable particle in the self-interactions of light.  We first observe that polarizability can arise either from the exchange of  mediator states or as a consequence of the inner structure of the particle. To exemplify this second possibility we calculate the polarizability of a neutral  bosonic open string, and find it is described only by dimension-8 operators.  Focussing on the spin-0 case, we calculate the  light-by-light scattering amplitudes induced by the dimension-6 and 8 polarizability operators.  
Performing a simulation of exclusive diphoton production with proton tagging at the LHC, we find that the imprint of the polarizable dark particle can be potentially detected at  $5\sigma$ significance
for  mass and cutoff reaching values above the TeV scale, for $\sqrt{s}=13$~TeV and $300$ fb$^{-1}$ of integrated luminosity.  If the polarizable dark particle is stable, it can be a dark matter candidate, in which case we argue this exclusive diphoton search may complement the existing LHC searches for polarizable dark matter.


\noindent
\clearpage

\section{Introduction}

\label{se:intro}

Among  the  speculations about what  lies beyond  the Standard Model (SM) of particles, there is the intriguing possibility  of   particles that are electrically neutral but can still slightly   interact with photons. 
The existence of such ``almost-dark'' particles is theoretically well-motivated. These could for example be   the  hadrons created by a hidden strongly-interacting gauge force, binding together  electrically charged constituents
(for recent scenarios featuring such bound states, see for instance  Stealth dark matter \cite{Appelquist:2014jch} and Vectorlike confinement \cite{Kilic:2009mi}).
Models where the dark particle has an electromagnetic coupling have been investigated in the scope of explaining Dark Matter (DM) of the universe. When interpreted as DM, the dark particle is assumed to be stable. In the present work this assumption of stability will not be needed.

In general, a particle with no electric charge may still interact with \textit{one} photon through a dipole operator and/or a charge radius operator - in which case the particle couples directly to the photon field strength $F^{\mu\nu}$.
Such scenarios for dark particles have been investigated in \cite{ 
Bagnasco:1993st,
Pospelov:2000bq,Sigurdson:2004zp,
Barger:2010gv,Banks:2010eh,Cho:2010br,An:2010kc,
Chang:2010en,McDermott:2010pa,DelNobile:2012tx,Vecchi:2013iza,
Rajaraman:2012db,Rajaraman:2012fu,Frandsen:2012db,D'Eramo:2014aba,Crivellin:2014qxa,Fedderke:2013pbc,Krall:2014dba} in the context of dark matter.
 However, if the neutral particle can be described  as a self-conjugate field  in a low-energy theory, these operators vanish.  \footnote{
By self-conjugate we mean a field transforming in a real representation: real scalar, real vector, Majorana fermion\ldots
}
 The main interaction of the dark particle with light  is then  controlled by its polarizability, \ie~its tendency to interact  with \textit{two} photons.
Such scenarios have also been investigated \cite{Pospelov:2000bq,Weiner:2012cb,Frandsen:2012db,Cotta:2012nj,Carpenter:2012rg,Crivellin:2014gpa,Ovanesyan:2014fha,Crivellin:2015wva,Appelquist:2015zfa,Brooke:2016vlw}, still in the context of dark matter.  
Such \textit{polarizable dark particles} are the topic of the present paper.

The interactions of our focus  are  bilinear in both  the dark particle and in $F^{\mu\nu}$. In the case of a scalar, a linear coupling of the form $\phi (F)^2$ may also exist in principle. We will assume this coupling is  either negligible or forbidden by a symmetry.\footnote{
The presence of a sizeable $\phi (F)^2$ term would considerably change 
the phenomenological prospects for the dark particle. In particular, the dark particle could be constrained by resonant production at colliders, and  by Casimir force experiments if its mass is below the keV scale. This scenario lies outside the scope of our study.
} By electroweak (EW) gauge invariance, a dark particle with electromagnetic polarizability should also be polarizable with respect to the  $W$ and $Z$ bosons.  This aspect will play little role in our analysis, but is relevant when it comes to   comparisons with the literature.

To the best of our knowledge of the literature, most of the searches for a dark particle polarizable by EW gauge bosons
are done within the assumption  that the dark particle is stable. This is true by definition for direct and indirect detection, and is also the case for collider searches \cite{Aad:2014tda,Aad:2014vka,Chatrchyan:2014tja,ATLAS:2014wra,Khachatryan:2014rwa,Khachatryan:2014rra,Khachatryan:2014tva,Cotta:2012nj,Carpenter:2012rg,Nelson:2013pqa,Lopez:2014qja,Crivellin:2015wva, Brooke:2016vlw}, where the search strategies always involve detection of large missing energy. 
If this kind of searches turned out to be successful, it would  provide a striking  signature for the existence of dark matter.

In this paper we would like to adopt a slightly different strategy. Instead of readily testing the existence of  a \textit{stable} dark particle, we propose to rather test the existence of a dark particle, whether it is stable or not. 
In such approach, the assessment of stability  is postponed to the post-discovery era, together with the characterization of the other properties of the new particle such as spin  and mass.
A consequence of this approach is that
   observing  a large missing energy  is not required anymore. Rather, one can set up a search which is independent  of the hypothesis of stability.  
Adopting this slightly different viewpoint naturally leads to consider searches for the effects of \textit{virtual} polarizable dark particles. 

From a theoretical viewpoint, one advantage of looking for virtual dark particles is that, if a full dark sector is present, all the dark polarizable particles contribute to the signal, and not only the stable ones. This implies that the signal is enhanced with respect to searches only focussed on stable dark particles (\textit{e.g.}~usual DM searches).
 In the following we will  consider the case of a single dark particle, unless stated otherwise.

As a first step, we will classify the CP-even polarizability operators up to dimension-8 and discuss their possible 
microscopic origin in Sec.~\ref{se:micro}. As an example, the intrinsic, dimension-8 polarizability of the neutral bosonic open string case is calculated in Sec.~\ref{se:string}. These preliminary studies are needed to establish under which conditions the virtual search we propose is relevant. The virtual process we will focus on in this paper is photon-photon scattering. The amplitudes in the spin-0  case are given in Sec.~\ref{se:cutoff}.  Moreover we will consider the exclusive channel, where outgoing protons remain intact and are detected.  The simulation and its results are presented in Sec.~\ref{se:LbL}, and Sec.~\ref{se:conclusions} contains our conclusions.

\section{Polarizability operators}
\label{se:micro}

We use a low-energy effective field theory (EFT) approach.  Here the set of CP-even polarizability operators up to dimension-8 is classified. One writes down the operators featuring two photon field strengths and two dark particles of a given spin. 
It will be claimed below that the dimension-6 operators can be vanishing depending on the UV origin of polarizability, hence the dimension-8  operators can potentially be the dominant ones. 
The cutoff scale is denoted $\Lambda$, and validity of the effective description of polarizability by local operators requires that the dark particle mass and the energy flowing through the polarizability vertices    be smaller than $\Lambda$ (see also Sec.~\ref{se:motivation}).

The effective Lagrangian describing the spin-$s$ polarizable dark particle has the form
\be
{\cal L}^s={\cal L}^s_{\rm kin} + \frac{}{}
{\cal L}^s_{6}+{\cal L}^s_{7}+{\cal L}^s_{8} +O\left(\Lambda^{-9}\right)\,. \label{eq:Leff}
\ee
with ${\cal L}^s_{4+n}=\sum_I c^s_I\,\Lambda^{-n}\,{\cal O}^s_I$.~\footnote{We use a metric with $(+,-,-,-)$  signature, except in Sec.~\ref{se:string}}.
One introduces the dual electromagnetic field strength $\tilde F^{\mu\nu}=\frac{1}{2}\epsilon^{\mu\nu \alpha \beta} F_{\alpha\beta}$, and one defines 
\be
(F)^2=F^{\mu\nu}F^{\mu\nu}\,,\quad (F.F)^{\mu\nu}=F^{\mu\rho}F^{\rho\nu}\,,\quad
(F\tilde{F})=F^{\mu\nu} \tilde F^{\mu\nu}\,,\quad
(F.\tilde{F})^{\mu\nu}=F^{\mu\rho} \tilde F^{\rho\nu}\,.
\ee
The coefficients of the operators of Eq.~\eqref{eq:Leff}  should be understood as given at the EFT matching scale $\Lambda$. These coefficients should be in general written as $c_i(\Lambda)$. However, only the coefficients defined at the $\Lambda$ scale will effectively appear in our results, hence we will simply refer to them as $c_i$ in the following.

The dark particle will be called $\phi$, $\psi$, $X^\mu$ for spin-$0$, $1/2$, $1$ respectively, and its mass will be  denoted $m$. 
The operators  allowed by EW gauge invariance and  inequivalent under field redefinitions  and integration by parts are classified in Tab.~\ref{tab:ops}. We do not include operators that induce a coupling to gauge bosons after EW breaking, which would arise from Higgs covariant derivatives like $ (D^\mu H)^\dagger D^\nu H$.
Also, the CP-even operators ${\cal O}^{0}_{8 \tilde{a}}=\partial^\mu \phi \partial^\nu \phi (\tilde{F}.\tilde{F})^{\mu\nu}$, ${\cal O}^{\nicefrac{1}{2}}_{8 \tilde{a}}=i\bar \Psi \gamma_\mu \partial^\nu \Psi  (\tilde{F}.\tilde{F})^{\mu\nu}$,
${\cal O}^{1}_{8 \tilde{a}}=X_\mu X_\nu (\tilde{F}.\tilde{F})^{\mu\nu}$  are not independent of the ones given in Tab.~\ref{tab:ops}, as they decompose as   
\be {\cal O}^{s}_{8 \tilde{a}} = 
{\cal O}^{s}_{8a} + 
\frac{1}{2}{\cal O}^{s}_{8b} \,,
\ee
and are thus not included.

\begin{table}[t]
\centering
\begin{tabular}{|c|c|c|c|c|}
\hline 
\multicolumn{4}{|c|}{Spin $0$} \\ 
\hline 
Operator & $\phi^2 (F)^2$ & $\partial^\mu \phi \partial^\nu \phi (F.F)^{\mu\nu}$  &  $(\partial^\mu \phi)^2 (F)^2$ \\ 
Dimension & 6 & 8 & 8  \\ Name & ${\cal O}^{0}_{6a}$ & ${\cal O}^{0}_{8a}$ & ${\cal O}^{0}_{8b}$  \\ 
\hline 
\end{tabular} 

\vspace{1cm}
\centering 
\begin{tabular}{|c|c|c|c|c|}
\hline 
\multicolumn{4}{|c|}{Spin $1/2$} \\ 
\hline 
Op. & $\bar \Psi \Psi (F)^2$ 
& $i\bar \Psi \gamma_\mu \partial^\nu \Psi  (F.F)^{\mu\nu}$
& $i\bar \Psi \gamma_\mu \partial^\mu \Psi  (F)^2$
 \\ 
 Dim. & 7  & 8 & 8   \\
 Name & ${\cal O}^{\nicefrac{ 1}{2}}_{7a}$ & ${\cal O}^{\nicefrac{ 1}{2}}_{8a}$ & ${\cal O}^{\nicefrac{ 1}{2}}_{8b}$    \\
\hline 
\multicolumn{1}{c|}{} 
& $i \bar \Psi \gamma_5 \Psi  (F\tilde{F})$ 
 & $\bar \Psi \gamma_5 \gamma_\mu   \partial^\nu \Psi   (F.\tilde F)^{\mu\nu} $   
& $\bar \Psi \gamma_5 \gamma_\mu   \partial^\mu \Psi   (F\tilde{F})$  \\
\multicolumn{1}{c|}{}  &  7
& 8 & 8  \\
\multicolumn{1}{c|}{} & ${\cal O}^{\nicefrac{ 1}{2}}_{7b}$ & ${\cal O}^{\nicefrac{ 1}{2}}_{8c}$
& ${\cal O}^{\nicefrac{ 1}{2}}_{8d}$ 
  \\
\cline{2-4}
\end{tabular} 
\vspace{1cm}

\centering 
\begin{tabular}{|c|c|c|c|c|c|}
\hline 
\multicolumn{6}{|c|}{Spin $1$} \\ 
\hline 
Op. & $(X_\mu)^2 (F)^2$ & $X_\mu X_\nu (F.F)^{\mu\nu}$ &     $(X.X)^{\mu\nu} (F.F)^{\mu\nu}$ 
& $(X)^2 (F)^2$
& $(X.F)^{\mu\nu} (X.F)^{\mu\nu}$   \\ 
Dim. & 6 & 6 & 8 &8 &8 \\ Name & ${\cal O}^{1}_{6a}$ & ${\cal O}^{1}_{6b}$ & ${\cal O}^{1}_{8a}$  & ${\cal O}^{1}_{8b}$ & ${\cal O}^{1}_{8c}$ \\ 
\hline 
\end{tabular} 

\caption{CP-even polarizability operators for a  self-conjugate particle of spin 0, 1/2 and 1.
The notations for  the field strength contractions of $X$ are the same as for $F$.
 \label{tab:ops} }
\end{table}

When truncating the EFT expansion at dimension 8, 
higher-order contributions to the coefficients of 
 operators with lower dimension should be kept up to dimension 8 (see~\cite{Contino:2016jqw} for related discussions). 
We will actually encounter such situation, with  dimension-6 and 7  operators coming respectively with a prefactor $\frac{m^2}{\Lambda^2}$ and $\frac{m}{\Lambda}$. 
These operators are  of dimension-8 in the sense they come with a $\Lambda^{-4}$ factor, and will be referred to as
\be
\hat{\O}^s_{ 6 i}\equiv\frac{m^2}{\Lambda^2} {\O}^s_{6i}\,, \quad \hat{\O}^s_{ 7 i}\equiv\frac{m}{\Lambda} {\O}^s_{7i}\,.
\label{eq:hatted}
\ee
The coefficients of the $\hat {\O}^s_{ 6 i}$ operators will be written $\hat c^s_{ 6 i}$, and similarly for $\hat {\O}^s_{ 7 i}$.

Finally, we remark that the dimension-6 and 7 operators can be naturally suppressed with respect to the dimension-8 ones if the dark particle has an approximate shift-symmetry. When this happens, the dark particle mass  should be suppressed similarly. One can parametrize the explicit breaking of the  shift-symmetry  using the dark particle mass, and the dimension-6 (-7) operators are then respectively suppressed by $m^2/\Lambda^2$, $m/\Lambda$ and can thus be identified as the hatted operators of Eq.\eqref{eq:hatted}.
This situation occurs  for instance if the dark particle is the Nambu-Goldstone particle of a spontaneously broken approximate global symmetry, for example a $U(n)$ symmetry or  supersymmetry,  respectively giving a Nambu-Goldstone scalar and a Nambu-Goldstini
 (see \cite{Bruggisser:2016nzw,Bruggisser:2016ixa} for a related analysis in the context of dark matter).

\subsection{Microscopic origin}

Even though we simply listed the polarizablity operators in an effective theory approach, some aspects of the UV origin of these operators can already be deduced. We identify two mechanisms. Either  polarizability  could arise from the exchange of  heavy  virtual particles, referred to as ``mediator''. Or the polarizable particle may actually be an extended object in the UV, and polarizability could then originate from the inner structure of the particle. 
  We shall refer to these two scenarios as \textit{mediated} polarizability and \textit{intrinsic} polarizability.

\subsubsection{Mediated polarizability} 

Here we consider the case of polarizability induced by heavy mediators.
First, we notice that no operator in Tab.~\ref{tab:ops}  can be generated via the tree-level exchange of a particle in a fully renormalizable theory. It may seem possible in the case of the ${\cal O}^1_{6b}$ operator, starting from a dipole operator  
\be \label{eq:XYF} {\cal O}^{1}_{XY}= X_\mu Y_\nu F^{\mu\nu}\,, \ee 
and integrating out the heavy spin-1 mediator $Y$. However, renormalisability requires $X$ and $Y$ to arise as massive gauge fields of a spontaneously broken gauge symmetry $G$ containing the electroweak group $G_{\rm EW}$ in its unbroken sector. 
This  ${\cal O}^{1}_{XY}$ operator can then arise from the kinetic term of the $G$ gauge field. Inspecting the broken sector (see \cite{Fichet:2013ola}), it turns out that ${\cal O}^{1}_{XY}$ is controlled by the broken constant structures $f^{\hat{a}\hat{b}c}$, where the hatted (unhatted) indexes label the broken (unbroken) generators. These same constant structures determine the coupling of the massive gauge fields to the electroweak gauge fields. One concludes that the $X$ and $Y$ fields have to be charged in order for ${\cal O}^{1}_{XY}$ to be non-zero. This is in contradiction with the hypothesis of a self-conjugate $X$, therefore polarizability of $X$ cannot be induced by tree-level exchange of a  spin-1  mediator in a renormalizable theory.

Possibilities for tree-level exchange of heavy mediators arise  in case of \textit{non-renormalizable} interactions.
The ${\cal O}^{0}_{6a}$, ${\cal O}^{0}_{8b}$, ${\cal O}^{\nicefrac{1}{2}}_{7a}$, ${\cal O}^{\nicefrac{1}{2}}_{8b}$, ${\cal O}^{1}_{6a}$, ${\cal O}^{1}_{8b}$ operators can be generated by a CP-even spin-0 mediator, such as a radion/dilaton.
The  ${\cal O}^{\nicefrac{1}{2}}_{7b}$, ${\cal O}^{\nicefrac{1}{2}}_{8d}$ could be induced by the exchange of an axion-like CP-odd scalar. 
The ${\cal O}^{\nicefrac{1}{2}}_{7a}$, ${\cal O}^{\nicefrac{1}{2}}_{7b}$ operators can also be generated together if a Majorana fermion $\Psi$ shares a dipole operator with another, heavier Majorana fermion $\Psi^*$, $\Psi\sigma_{\mu\nu}\Psi^* F^{\mu\nu}$ \cite{Weiner:2012cb}.  A similar possibility is that the components of $\Psi$ and $\Psi^*$ be part of a single Dirac fermion, with a mass splitting induced by a Majorana mass \cite{DeSimone:2010tf}.  
Finally, the ${\cal O}^{0}_{8a}$, ${\cal O}^{\nicefrac{1}{2}}_{8a}$, ${\cal O}^{1}_{8a}$ can be generated by a spin-2 mediator (such as a Kaluza-Klein graviton), together with ${\cal O}^{0}_{8b}$, ${\cal O}^{\nicefrac{1}{2}}_{8b}$, ${\cal O}^{1}_{8b}$  terms coming from tracelessness of the spin-2 representation.

All of the operators of Tab.~\ref{tab:ops} could in principle be generated at loop-level, in particular by loops of charged mediators. In such case, the coefficient $c^s_i$ of the polarizability operators must come with a factor $e^2/16\pi^2$, and $\Lambda$ is  identified with the mass of the particle in the loop. 

\subsubsection{Intrinsic polarizability} 

Here, we consider the possibility that polarizability arises from the inner structure of the dark particle.
Let us consider a generic 4-point amplitude with two dark particles and two photons in external legs. We focus on the scalar case $\gamma\gamma\phi\phi$ for concreteness, but the same reasoning applies to spin 1/2 and 1 similarly. 
 The scattering amplitude has the form
\be
{\cal M}=\epsilon_\alpha(p_a)\epsilon_\beta(p_b ) V^{\alpha\beta}\,,
\ee
where  $V^{\mu\nu}$ is a function of the momenta and of  the intrinsic scale of the dark particle $\Lambda$. Using Ward identities and  the fact that the photon does not couple to the dark particle through covariant derivatives by definition, we readily know that the $V^{\alpha\beta}$ tensor has the form
\be
V^{\alpha\beta}=\frac{1}{m^2} R^{\alpha \mu_a \nu_a}(p_a)R^{\beta \mu_b \nu_b}(p_b)\,F^{\mu_a \nu_a \mu_b \nu_b} \label{eq:V}
\ee
where one introduces the projector $R^{\alpha \mu \nu}(p)=p^\mu g^{\alpha \nu}-p^\nu g^{\alpha \mu}$. The dimensionless tensor $F^{\mu_a \nu_a \mu_b \nu_b}$ is the general form factor of the dark particle, that encodes the information about its inner structure.  In the low-energy domain $s,t,u,m^2<\Lambda^2$, where $s,t,u$ are the Mandelstam variables, the lower order Lorentz structures can then be written as 
\be
F^{\mu_a \nu_a \mu_b \nu_b}=F_0(s,t,u,\Lambda) g^{\mu_a\mu_b}g^{\nu_a\nu_b}+ \frac{1}{\Lambda^2} F_1(s,t,u,\Lambda) p_1^{\mu_a}p_2^{\mu_b}g^{\nu_a\nu_b}+O\left(\frac{1}{\Lambda^4}\right)\,.
\ee
We assume that a massless polarizable dark particle can exist, and thus  ask for the amplitude to remain finite in the massless limit. This implies that the form factors should decrease at least as $F_{0,1}\sim m^2/\Lambda^2$ at small $m/\Lambda$ in order to compensate the $m^{-2}$ in Eq.~\eqref{eq:V}.  This can also be checked taking the massless limit of the amplitudes of Sec.~\ref{se:amp}.   Expanding  the form factors for large $\Lambda$ and using the symmetries of the diagram~\footnote{Because of $t\leftrightarrow u$ symmetry one can expand with respect to $s/\Lambda^2$ and $(t+u)/\Lambda^2$. One uses then $s+t+u=2m^2$.  }, one gets that the leading terms should be given by $F_0=\frac{m^2}{\Lambda^2}\tilde F_0$, $F_1=\frac{m^2}{\Lambda^2}\tilde F_1$ and   $\tilde F_0(s,t,u,\Lambda)=A+ (B p_1.p_2+ C m^2 ) /\Lambda^2+O(\Lambda^{-4})$, $\tilde F_1(s,t,u,\Lambda)=D+O(\Lambda^{-2})$ where $A$, $B$, $C$, $D$ are  constants. The general form factor reads
\be
F^{\mu_a \nu_a \mu_b \nu_b}=\frac{m^2}{\Lambda^2}(A+\frac{B p_1.p_2+C m^2}{\Lambda^2}) g^{\mu_a\mu_b}g^{\nu_a\nu_b}+D \frac{m^2}{\Lambda^4} p_1^{\mu_a}p_2^{\mu_b}g^{\nu_a\nu_b}+O\left(\frac{1}{\Lambda^6}\right)\,.
\ee
All the terms vanish in the pointlike limit $\Lambda\rightarrow \infty$,  as expected from  effects arising from compositeness.
%
The $A$, $B$, $C$, $D$ constants are in direct correspondence with the spin-0 effective operators of Tab.~\ref{tab:ops}. Identifying the Lorentz structures, one has simply 
\be
A = c_{6a}^0\,\quad B = c_{8b}^0 \,\quad C = \hat c_{6a}^0\,\quad D = c_{8a}^0\,.
\ee

We can deduce some physical features of the polarizability operators by studying the non-relativistic limit $(p_i)^2\ll m^2$, where $p_i$ is the three-momentum. Here we limit our discussion to spin-0 and $\nicefrac{1}{2}$, for which one can recognize familiar electromagnetic features.~\footnote{The case of a massive non-relativistic  spin-1 particle is not straigthforward to analyze, for example one has ${\cal O}_{8a}^1\propto X_iX_j\,({ E}_i{ E}_j+{ B}_i{ B}_j - \delta_{ij}|{ B}|^2. )$ }
Let us first remark that in the non-relativistic limit the operators satisfy
\be
{\cal O}_{8a}^0\propto  (E_i)^2 \,, \quad {\cal O}_{6a}^0,\, {\cal O}_{8b}^0 \propto (E_i)^2-(B_i)^2\,, 
\ee
\be
{\cal O}_{8a}^{\nicefrac{1}{2}}\propto  (E_i)^2 \,, \quad  {\cal O}_{7a}^{\nicefrac{1}{2}},\,{\cal O}_{8b}^{\nicefrac{1}{2}} \propto (E_i)^2-(B_i)^2\,, \quad
{\cal O}_{7,b}^{\nicefrac{1}{2}}\,,{\cal O}_{8c,d}^{\nicefrac{1}{2}}\rightarrow  0 \,,
\ee
where $E_i$, $B_i$ are the standard electric and magnetic fields. 
One also has
\be
 {\cal O}_{8 \tilde{a}}^0\propto  (B_i)^2 \,,  \quad  {\cal O}_{8 \tilde{a}}^{\nicefrac{1}{2}}\propto  (B_i)^2 \,. 
\ee
The $(E_i)^2$ and $(B_i)^2$ term that appear in the non-relativistic Lagrangian  correspond respectively to the static electric and magnetic susceptibilities of the inner structure of the dark particle.
We can see that the ${\cal O}_{8a}^{0,\nicefrac{1}{2}}$ (${\cal O}_{8 \tilde{a}}^{0,\nicefrac{1}{2}}$) operators describe respectively a polarizability with purely electric (respectively magnetic) origin. These properties can in turn be used to infer some features of the polarizability operators for a given object.

In the case of dark hadrons, made of electrically charged fermions glued by a hidden strong interaction, we certainly expect an electric polarizability, as these constituents form an electronic density that can be deformed by an external electric field.  Also, as the constituents carry intrinsic spin, a magnetic  polarizability should exist, however both theoretical arguments \cite{Luke:1992tm} and observations \cite{Agashe:2014kda} suggest that it is suppressed with respect to the electric one, thus one may expect $c_{8 b}^0+\hat c_{6a}^0+ \frac{\Lambda^2}{m^2}c_{6a}^0\ll c_{8a}^0$, $c_{8 b}^{\nicefrac{1}{2}}+\hat c_{6a}^{\nicefrac{1}{2}}+ \frac{\Lambda}{m}c_{6a}^{\nicefrac{1}{2}}\ll c_{8a}^{\nicefrac{1}{2}}$. 

The case of a neutral string with non-zero charges at endpoints 
is also interesting. 
 In that case one should consider the effective operators associated with the quantum states of the string. 
\footnote{The dipole operators associated with the quantum states of the open bosonic and super strings have been evaluated in Ref.~\cite{Ferrara:1992yc} and are vanishing if the string is neutral.} An electric polarizability should exist as one has two charges binded together. In contrast, as no intrinsic spin is attached to any point of the string, the string cannot have a magnetic polarizability. We thus expect $c_{8a}^0\neq 0$, $c_{8 b}^0+\hat c_{6a}^0+ \frac{\Lambda^2}{m^2}c_{6a}^0=0$, and simlarly for spin-$1/2$.  These coefficients will be calculated in next section for the neutral bosonic string. 

Finally, one may wonder how electro-magnetic duality applies to the arguments above. From a macroscopic viewpoint, electromagnetic duality exchange the susceptibilities $\alpha_E\leftrightarrow \alpha_B$, and thus exchanges the $\alpha_E\neq 0,\alpha_M=0$ case with $\alpha_E= 0,\alpha_M \neq 0$ in the string case.
Microscopically, such object would  be a sort of open string with magnetic monopoles attached at endpoints. Such objects, called $D$-strings, do exist in string theories, and are related by S-duality to the original strings (see \cite{becker2006string}). From a low energy point of view, the polarizability of such objects 
should  be expected to be described by the ${\cal O}_{8 \tilde{a}}$ operator, while the combinations $c_{8 b}^0+\hat c_{6a}^0+ \frac{\Lambda^2}{m^2}c_{6a}^0$ ($c_{8 b}^{\nicefrac{1}{2}}+\hat c_{6a}^{\nicefrac{1}{2}}+ \frac{\Lambda}{m}c_{6a}^{\nicefrac{1}{2}}$) should again vanish.

 \section{Polarizability of the neutral bosonic string }
\label{se:string}

To give a concrete example of an object with intrinsic polarizability,  we work out the case of a neutral open string (\ie~a string with charges $q_0$, $q_1$ at ends, satisfying $q_0=-q_1=q$).   For the sake of describing polarizability of the string states, there is no need to assume that spacetime has critical dimension. In fact, being ultimately interested in the 4D case, the string we consider cannot be considered as a fundamental one. Instead, it may for example be taken as a QCD-like string, \ie~an effective description of the binding between a quark and an antiquark  arising in a gauge theory with large number of colors.  
  A mostly-plus signature $(-,+,+,+)$  is used for $g^{\mu\nu}$ in this section.

The action of an open string with length scale $l_s\equiv\sqrt{\alpha'}$ in an electromagnetic background is given by 
\be
S=\int_{-\infty}^{\infty} d\tau \int_{0}^{\pi} d\sigma \left[ \frac{1}{4\pi l_s^2} \left({\dot X}^\mu {\dot X}_\mu-{\dot X}^{'\mu} X^{'}_{\mu} \right)- A^{\mu}{\dot X}_\mu\Big(q_0 \delta(\sigma)+q_1 \delta(\sigma-\pi) \Big)\right]\,,
\label{eq:S_string}
\ee
where $A_\mu$ is the canonically normalized electromagnetic field.
Propagation of a bosonic open string in an abelian background gauge field has been worked out in Ref.~\cite{Burgess:1986dw}, and canonical quantization is done in details in Ref.~\cite{Abouelsaood:1986gd}. Our calculation  follows closely \cite{Abouelsaood:1986gd}, details are given in App.\ref{app:string}. 

Computing the  solutions of the equation of motion, defining an orthogonal basis for the oscillator and zero modes, and asking for canonical commutators between the position, momentum and Fourier operators $x_\mu, p_\mu, a^{(\dagger)}_n$,  the string decomposition over orthonormal modes is 
\be\begin{split}
X_\mu= x_\mu + 2l^2_s\,\Big( g - 4\pi^2 l_s^4\,q^2\, F.F \Big)^{-1/2}_{\mu\nu} \Big(\tau\,g^{\mu\rho} + 2\pi l^2_s\,\left(\sigma-\frac{\pi}{2}\right)\,q\, F^{\nu\rho}     \Big) p_\rho
\\+i\sqrt{2}l_s\,\left(\sum_{n=1}^{\infty}a_n \psi_n(\tau,\sigma)- \sum_{n=1}^{\infty}a^\dagger_n \psi_{-n}(\tau,\sigma)   \right)\,.\end{split}
\label{eq:string_sol}
\ee
It turns out that the background field does not affect the oscillator modes, only the zero mode gets deformed.
\footnote{ There is a freedom in normalising the $x^\mu$ and $p^\mu$ operators inside the zero mode. 
 It is convenient to let the position operator unchanged and to incorporate all the effect of the background field into the momentum term. }
The $L_0$ Virasoro operator of the open string is then given by 
\be
L_0= \frac{1}{2}p^\mu \Big(g- (2\pi)^2\, l_s^4\,q^2\, F.F\,\Big)^{-1}_{\mu\nu}\, p^\nu + \frac{1}{2}N \label{eq:L0}
\ee
where  $N=\sum_{n=1}^{\infty} \,\alpha^\mu_{-n} \alpha_{n\,\mu}$ is the usual number operator, using $\alpha_{n}=\sqrt{n}\,a_n$, $\alpha_{-n}=\sqrt{n}\,a_n^{\dagger}$.
The states of the string are built from a ground state $\left|0\right\rangle$ using creation operators, \be
\Phi^{\mu_1\mu_1\ldots\mu_s}(x^\mu)= \int \frac{dk^4}{(2\pi)^4 }\, e^{ik_\mu x^\mu}\, \prod_{i=1}^s\, \alpha_{-m_i}^{\mu_i} \, \left|0\right\rangle\,.
\ee
The $L_0$ operator satisfies the condition 
\be(L_0+a)\Phi=0 \,, \label{eq:string_eqL0} \ee
where $a$ is a constant from normal ordering which is left unspecified and is irrelevant regarding the property of polarizability. \footnote{We will assume $a\geq 0$ whenever discussing the spin-0 state, otherwise it is tachyonic. }
Equation \eqref{eq:string_eqL0}
 gives the equation of motion for the string states, 
 \be
\left(\partial^{\mu} \Big(g- (2\pi)^2\, l_s^4\,q^2\, F.F\,\Big)^{-1}_{\mu\nu}\,\partial^{\nu} -m^2 \right)\Phi^{\mu_1\,\mu_2\ldots \mu_s}=0 \,,\label{eq:string_EOM}
\ee
where the mass is given by $m=  l_s^{-1}(\sum_i m_i+a)^{1/2}$.
Retaining the leading term  in  power of $l_s$ gives
\be
\bigg(\partial_{\mu} \left( g_{\mu\nu} + (2\pi)^2\, l_s^4\,q^2\, (F.F)_{\mu\nu} +O(l_s^8) \bigg) \partial_{\nu} -m^2 \right)\Phi^{\mu_1\,\mu_2\ldots \mu_s}=0 \,. \label{eq:string_EOM_exp}
\ee
This equation of motion describes the polarizability of a string state of any integer spin $s$.
Going back to the mostly-minus metric used for the effective Lagrangian of Eq.~\eqref{eq:Leff}, we can deduce the  Lagrangian giving rise to the equation of motion Eq.~\eqref{eq:string_EOM_exp} in case of spin $0$ and $1$.
We conclude that  polarizability of the spin-0 state and spin-1 state is respectively described by the operators $\O^{0}_{8a}$, $\O^{1}_{8a}$.
Identifying $\Lambda$ with the inverse string length, $\Lambda=l_s^{-1}$,  the operator coefficients are $(2\pi)^2\,q^2$, so that the effective Lagrangian is
\be
{\cal L}\supset \frac{4\pi^2\, q^2}{\Lambda^4} \O^{0}_{8a} +  \frac{4\pi^2\,q^2}{\Lambda^4} \O^{1}_{8a}\,. \label{eq:Opstring}
\ee

 Establishing the consistent Lagrangian for a neutral polarizable state of higher spin is probably more  challenging conceptually and technically, and lies outside the scope of this study.  In particular,  the electromagnetic interactions of the auxiliary fields present in the higher-spin Lagrangian would have to be determined.~\footnote{These aspects might be treated in a further work.}

 \section{Four-photon amplitudes  from polarizable dark particles}
\label{se:cutoff}

\begin{figure}
\centering
\begin{picture}(170,100)
\put(0,0){
\put(0,0){\includegraphics[scale=0.6,clip=true, trim= 0cm 0cm 0cm 0cm]{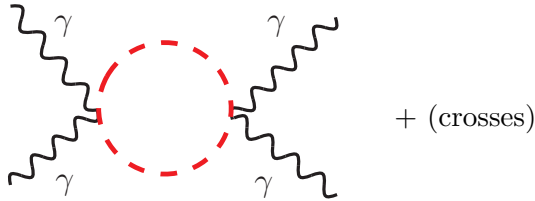}}
\put(150,30){\rotatebox{0}{$+$ (crosses) }}
}
\end{picture}
\caption{Four-photon interaction induced by a virtual polarizable dark particle.
\label{fig:loop}
}
\end{figure}

Polarizable dark particles automatically induce loops with four external photon legs (see Fig.~\ref{fig:loop}).
Following our strategy of focussing on virtual processes (see Sec.~\ref{se:intro}), 
we propose to use such  anomalous photon couplings as a probe for the existence of a dark particle. 
For a first analysis of this proposal,  we focus on the case of a dark particle of spin-0. The spin-1/2 and spin-1 cases would deserve to be treated similarly, but lie outside  the scope of this paper.

\subsection{Consistency of the approach}
\label{se:motivation}

A necessary condition for our proposal to make sense is that the dark particle produces the main contribution to the four-photon coupling. While in principle  the complete UV picture is needed to answer this question,
the considerations on the microscopic origin of polarizability made in Sec.~\ref{se:micro} already provide a useful constraint. 
 Indeed, in the case where polarizability is induced by  mediators, the mediators themselves  can form diagrams with four external photons. 
The contributions from dark particles are expected to be smaller than the ones from mediators by at least a loop factor. This happens in both the cases of loop and tree diagrams, induced respectively by  charged mediators and mediators with non-renormalizable couplings. The four-photon search then essentially probes the existence of these mediators. The sensitivity for such particles has already been estimated \cite{Fichet:2013gsa,Fichet:2014uka}, irrespective of the existence of a dark particle. 
In contrast, if polarizability originates from the inner structure of the dark particle, the dark particle loop can in principle be the dominant contribution to the anomalous four-photon vertex. 

Some consistency constraints also come from the validity of the EFT approach. The validity of the low-energy expansion requires  that \be s,|t|,|u|,m^2 < \Lambda^2\,,\ee otherwise the form factor from UV physics becomes important, and the description of polarizability of the dark particle by local operators is not valid anymore. The  partonic center-of-mass energy for exclusive photon scattering is typically of $\sqrt{s_{\gamma\gamma}}\sim 1$~TeV at the 13 TeV LHC.
Moreover, tree-level unitary of photon\,-\,dark particle scattering imposes the conditions
\be
|c^0_{8a}| s^2 /\Lambda^4<16 \pi\,,\quad
|c^0_{8b}| s^2 /\Lambda^4<8 \pi\,,\quad
|c^0_{6a}| s /\Lambda^2<8 \pi\,, \quad
|\hat c^0_{6a}| m^2s /\Lambda^4<8 \pi\,. \label{eq:unit_bound}
\ee
For $\sqrt{s}\sim \Lambda$, the bound translates as $c_i<8\pi$. It is worth noticing that for  $\Lambda>\sqrt{s}$, the $c_i$ are allowed to be larger than $8\pi$.  In our  estimations of LHC sensitivity of Sec.~\ref{se:LbL} we will use $|c^0_i|=10$. We emphasize that these unitarity constraints are qualitatively equivalent to requiring  perturbativity of the effective interactions in the EFT. Constraints similar to those of Eq.~\eqref{eq:unit_bound} can be obtained by requiring that a diagram with $n+1$ loops be smaller or of same order of magnitude than a diagram with $n$ loops (when using dimensional regularization).

\subsection{Consistency of the calculation}
\label{se:consistency_cal}

An important subtlety is that the four-photon loop diagrams we consider come from higher-dimensional operators and are thus more divergent than the four-photon diagrams from the UV theory. 
This implies that four-photon \textit{local} operators (\ie~counter-terms) are also present in the effective Lagrangian to cancel the divergences which are not present in the UV theory. 
 The finite contribution from these local operators is fixed by the UV theory at the matching scale, 
  and is expected to be of same order as the coefficient of the $\log \Lambda$ term in the amplitude by naive dimensional analysis   (this situation is  analog to renormalisation of the non-linear sigma model, see Ref.~\cite{Manohar:1996cq}). This implies that the amplitudes obtained from calculating the loop graphs should only be considered as \textit{estimates} of the complete amplitudes, the latter being determined only once the UV theory is specified. 
Concretely, for four-photon interactions induced by loops with dimension-8 operators, local four-photon operators of dimension-12 are present in the Lagrangian. 
Four-photon interactions induced by loops of dimension-6 operators imply the presence of dimension-8 operators,  corresponding to the two Lorentz structures shown in Eq.~\ref{eq:dim8loc}.

Cutoff regularisation in an effective theory is very difficult  because it breaks the expansion with respect to $\Lambda^{-1}$, as loops from operators of arbitrarily high dimension contribute at same order to the amplitudes (see \cite{Manohar:1996cq}). A much simpler scheme is dimensional regularisation, in which case power-counting is respected and it is thus consistent to include only operators of lower dimension (up to dimension-8 in our case).  The matching of the effective theory with the UV theory being done at the scale $\Lambda$, we can readily identify the divergent integrals as  (see \cite{peskin1995introduction,Alam:1997nk})~\footnote{The running of the $c^0_i(\mu)$ coefficients is taken into account at leading-log order with this method. }
\be
\int \frac{d^4 l}{(2\pi)^4}\frac{1}{(l^2-\Delta)^2} \rightarrow \frac{-i}{(4\pi)^2}\log(\Delta/ \Lambda^2)\,,
\ee
\be
\int \frac{d^4 l}{(2\pi)^4}\frac{l^2}{(l^2-\Delta)^2} \rightarrow \frac{-2\,i }{(4\pi)^2}\Delta\log( \Delta/ \Lambda^2)\,,
\ee
\be
\int \frac{d^4 l}{(2\pi)^4}\frac{(l^2)^2}{(l^2-\Delta)^2} \rightarrow \frac{-3\,i }{(4\pi)^2}\Delta^2\log( \Delta/ \Lambda^2)\,.
\ee

As a final remark, we note that in the limit of heavy mass, $m^2 \gg s,t,u$, the loops reduce to local effective interactions. The amplitudes from these local interactions are given in \cite{Fichet:2014uka}, and have the Lorentz structure
\be
{\cal M}_{++++} \propto s^2\,,\quad {\cal M}_{++--} \propto s^2+t^2+u^2\,,\label{eq:ampEFT}
\ee 
and ${\cal M}_{+++-}=0$ (see next subsection for definition of helicity states). As a mild sanity check for our loop calculations, we observe that all our amplitudes reproduce the structure of Eq.~\ref{eq:ampEFT} at first order in the $ O(s,t,u/m^2)$ expansion. The coefficients of the local dimension-8 operators corresponding to each loop can  also be deduced from Eq.~\eqref{eq:ampEFT}, and will be given below.

\subsection{Helicity amplitudes}
\label{se:amp}

Focussing on the case of a spin-0 dark particle, we calculate the four-photon amplitudes induced by the dimension-8 polarizability operators $\O^0_{8a}$, $\O^0_{8b}$, $\hat \O^0_{6a}$, which are theoretically well-motivated as discussed in Sec.~\ref{se:motivation}.  We limit ourselves to cases where one of these operators is dominant 
and do not calculate diagrams involving two different operators.

Helicity amplitudes are given under the form ${\cal M}_{\lambda_a\lambda_b\lambda_1\lambda_2}(s,t,u)$, where $\lambda_{a,b}=\pm$ denotes the polarization of two ingoing photons and $\lambda_{1,2}$ denotes the polarization of two outgoing photons.
Due to the relations ${\cal M}_{+-+-}(s,t,u)={\cal M}_{++++}(u,t,s)$, ${\cal M}_{+--+}(s,t,u)={\cal M}_{++++}(t,s,u)$, only the ${\cal M _{++++}}$, ${\cal M _{++--}}$, ${\cal M _{+++-}}$ configurations have to be calculated (see Ref.~\cite{Costantini:1971cj}).
Full amplitudes and details of the calculation are given in App.~\ref{app:LbL}. The ${\cal M}_{+++-}$ amplitude is found to be exactly zero in all cases.  Here below we display only the helicity amplitudes in the high energy limit $m^2 \ll s,t,u  $ and in the low-energy limit $s,t,u\ll m^2$ , where in both cases $s,t,u,m^2< \Lambda^2$. 
\begin{itemize}
\item $\O^0_{8a}$ operator

If $m^2 \ll s,t,u  $, 
\be\begin{split}
{\cal M}_{++++}\approx -\frac{(c^0_{8a})^2}{32 \pi^2\,\Lambda^8} \,s^2 \bigg[ -\left(\frac{68}{75} s^2 +\frac{47}{900} (t^2+u^2) \right)+i\pi\left(\frac{3}{5}s^2+\frac{1}{30}(t^2+u^2)\right)+ \\ \left(\frac{3}{5}s^2\log\left(\frac{s}{\Lambda^2}\right)+\frac{1}{30}(t^2\log\left(\frac{t}{\Lambda^2}\right)+u^2\log\left(\frac{u}{\Lambda^2}\right))\right) \bigg]\,,
\end{split}
\ee
\be\begin{split}
{\cal M}_{++--}\approx -\frac{(c^0_{8a})^2}{32 \pi^2\,\Lambda^8} \bigg[& \left(\frac{-68}{75} +i\frac{3 \pi}{5}\right)(s^4+t^4+u^4) + \\ & \frac{3}{5}\left(s^4\log\left(\frac{s}{\Lambda^2}\right)+t^4\log\left(\frac{t}{\Lambda^2}\right)+u^4\log\left(\frac{u}{\Lambda^2}\right) \right)\bigg]\,.
\end{split}
\ee

If $m^2 \gg s,t,u  $, 
\be
{\cal M}_{++++}\approx -\frac{(c^0_{8a})^2}{32 \pi^2\,\Lambda^8} \,5\, s^2 m^4 \log\left(\frac{m^2}{\Lambda^2}\right) \,,\label{eq:LE8app}
\ee 
\be
{\cal M}_{++--}\approx -\frac{(c^0_{8a})^2}{32 \pi^2\,\Lambda^8} \,3 \, (s^2+t^2+u^2 ) m^4 \log\left(\frac{m^2}{\Lambda^2}\right) \,.  \label{eq:LE8amm}
\ee

\item $\O^0_{8b }$ operator

If $m^2 \ll s,t,u  $, 

\be\begin{split}
{\cal M}_{++++}= -\frac{(c^0_{8b})^2}{8 \pi^2\,\Lambda^8} s^4 \bigg[ -\frac{157}{225} +i\pi\frac{7}{15}  + \frac{7}{15}\log\left(\frac{s}{\Lambda^2}\right) \bigg] \,,
\end{split}
\ee
\be\begin{split}
{\cal M}_{++--}= -\frac{(c^0_{8b})^2}{8 \pi^2\,\Lambda^8} \bigg[& \left(\frac{-157}{225} +i\pi\frac{7 }{15}\right)(s^4+t^4+u^4) + \\ & \frac{7}{15}\left(s^4\log\left(\frac{s}{\Lambda^2}\right)+t^4\log\left(\frac{t}{\Lambda^2}\right)+u^4\log\left(\frac{u}{\Lambda^2}\right) \right)\bigg]\,.
\end{split}
\ee

If $m^2 \gg s,t,u  $, 

\be
{\cal M}_{++++}= -3\frac{(c^0_{8b})^2}{2 \pi^2\,\Lambda^8} \, s^2 m^4 \log\left(\frac{m^2}{\Lambda^2}\right) \,,\label{eq:LE8bpp}
\ee 
\be
{\cal M}_{++--}= -3\frac{(c^0_{8b})^2}{2 \pi^2\,\Lambda^8} \, (s^2+t^2+u^2) m^4 \log\left(\frac{m^2}{\Lambda^2}\right) \,. \label{eq:LE8bmm}
\ee

\item $\hat\O^0_{6a }$ operator

If $m^2 \ll s,t,u  $, 

\be\begin{split}
{\cal M}_{++++}= -\frac{(\hat c^0_{6a})^2}{2 \pi^2\,\Lambda^8} \,m^4\, s^2 \bigg[ -2 +i\pi  + \log\left(\frac{s}{\Lambda^2}\right) \bigg] \,,
\end{split}
\ee
\be\begin{split}
{\cal M}_{++--}= -\frac{(\hat c^0_{6a})^2}{8 \pi^2\,\Lambda^8} m^4 \bigg[& \left(-2 +i\pi\right)(s^2+t^2+u^2) + \\ & \left(s^2\log\left(\frac{s}{\Lambda^2}\right)+t^2\log\left(\frac{t}{\Lambda^2}\right)+u^2\log\left(\frac{u}{\Lambda^2}\right) \right)\bigg]\,.
\end{split}
\ee

If $m^2 \gg s,t,u  $, 

\be
{\cal M}_{++++}= -\frac{(\hat c^0_{6a})^2}{2 \pi^2\,\Lambda^8} \, s^2 m^4 \log\left(\frac{m^2}{\Lambda^2}\right) \,,\label{eq:LE6ahatpp}
\ee 
\be
{\cal M}_{++--}= -\frac{(\hat c^0_{6a})^2}{2 \pi^2\,\Lambda^8} \, (s^2+t^2+u^2) m^4 \log\left(\frac{m^2}{\Lambda^2}\right) \,. \label{eq:LE6ahatmm}
\ee

\end{itemize}
Finally, in the $m^2 \gg s,t,u  $ case, it is well-known that four-photon interactions can be  represented by two independent dimension-8 operators 
\be
{\cal L} = \frac{b_1}{\Lambda^4} F^{\mu\nu}F_{\mu\nu} F^{\rho\sigma}F_{\rho\sigma} + \frac{b_2}{\Lambda^4} F^{\mu\nu} F_{\nu\rho}F^{\rho\sigma}F_{\sigma\mu}\,, \label{eq:dim8loc}
\ee
and the helicity amplitudes as a function of the $b_{1,2}$ coefficients have been given in Ref.~\cite{Fichet:2013gsa}.~\footnote{To adapt the amplitudes given in the conventions of \cite{Fichet:2013gsa} to the ones of the present paper, the amplitudes in \cite{Fichet:2013gsa} have to be multiplied by a factor $-8$.
} Matching these amplitudes to the low-energy limit of the ones from loops of polarizable particles Eqs.~\eqref{eq:LE8app},~\eqref{eq:LE8amm},~\eqref{eq:LE8bpp},~\eqref{eq:LE8bmm},~\eqref{eq:LE6ahatpp},~\eqref{eq:LE6ahatmm}  gives 
\be
b_1= - \frac{(c^0_{8a})^2}{64 \pi^2\,\Lambda^4}  m^4 \log\left(\frac{m^2}{\Lambda^2}\right)\,,\quad b_2=  -\frac{(c^0_{8a})^2}{128 \pi^2\,\Lambda^4}  m^4 \log\left(\frac{m^2}{\Lambda^2}\right)\,.
\ee
from the $\O^0_{8a}$ operator, 
\be
b_1= - 3 \frac{(c^0_{8b})^2}{16 \pi^2\,\Lambda^4} \,  m^4 \log\left(\frac{m^2}{\Lambda^2}\right)\,,\quad b_2=0
\ee
from the  $\O^0_{8b}$ and
\be
b_1=  - \frac{(\hat c^0_{6a})^2}{16 \pi^2\,\Lambda^4} \,  m^4 \log\left(\frac{m^2}{\Lambda^2}\right)\,,\quad b_2=0
\ee
from the $\hat\O^0_{6a}$ operator.

 \section{Light-by-light scattering as a probe for polarizable dark particles}
\label{se:LbL}

We propose to focus on photon-photon scattering in the exclusive channel, where the two protons remain intact after the collision,
\be
pp\rightarrow \gamma\gamma \,p p \,.
\ee
 These intact protons can be detected and characterised using  forward proton detectors along the beam pipe, that are scheduled by both ATLAS \cite{atlas} and  CMS/TOTEM \cite{cms} collaborations. 
The interest of the exclusive diphoton channel with proton characterization is that there is enough kinematic information to eliminate most of the background.
The sensitivity of this measurement to new physics has been studied in details in \cite{Fichet:2013gsa,Fichet:2014uka,Fichet:2016prl}, where the residual background rate after all cuts has been estimated to $3\cdot 10^{-4}$~fb.
This background  comes from  inclusive diphoton events occuring simultaneously with the tagging of two intact protons from pileup. 
~\footnote{ Other studies using proton-tagging at the LHC 
for New Physics searches can be found in 
Refs.~\cite{usww, usw, Gupta:2011be,Fichet:2013ola,Sun:2014qoa,
Sun:2014qba,Sun:2014ppa,Sahin:2014dua,Inan:2014mua,YellowReport}.
We refer to \cite{friend:2013yra} for a study of light-by-light scattering at the LHC without proton tagging.}

\subsection{Sensitivity  at $13$ TeV and $L=300$ fb$^{-1}$}

In order to obtain a realistic estimation for the discovery potential of the dark particle, we implemented the four-photon amplitudes induced by dark particles in the Forward Physics Monte Carlo generator (\texttt{FPMC} \cite{Boonekamp:2011ky}).
The model of photon flux of Ref.~\cite{Budnev:1974de} is assumed. 
We reproduce  the acceptance of the forward detectors by constraining the fractional momentum loss of both protons to be~\footnote{For CMS, these expectations have recently been updated to be $0.037<\xi<0.15$ \cite{private}.
 We checked that our results are essentially the same with this new range, the sensitivity regions decrease only slightly. }
\be
0.015<\xi<0.15\,.
\ee
We set a cut of \be |p_T|>150~{\rm GeV} \ee
 on the transverse momentum of each photon. Like in Ref.~\cite{Fichet:2014uka}, the main impact on the signal rates is  expected to come from these cuts. We include the effect of the other cuts on the signal with a global  efficiency of $\epsilon_s=90\%$.

\begin{figure}[t]
\centering
\begin{picture}(500,200)
\put(0,0){\includegraphics[scale=0.6,clip=true, trim= 0cm 0cm 0cm 0cm]{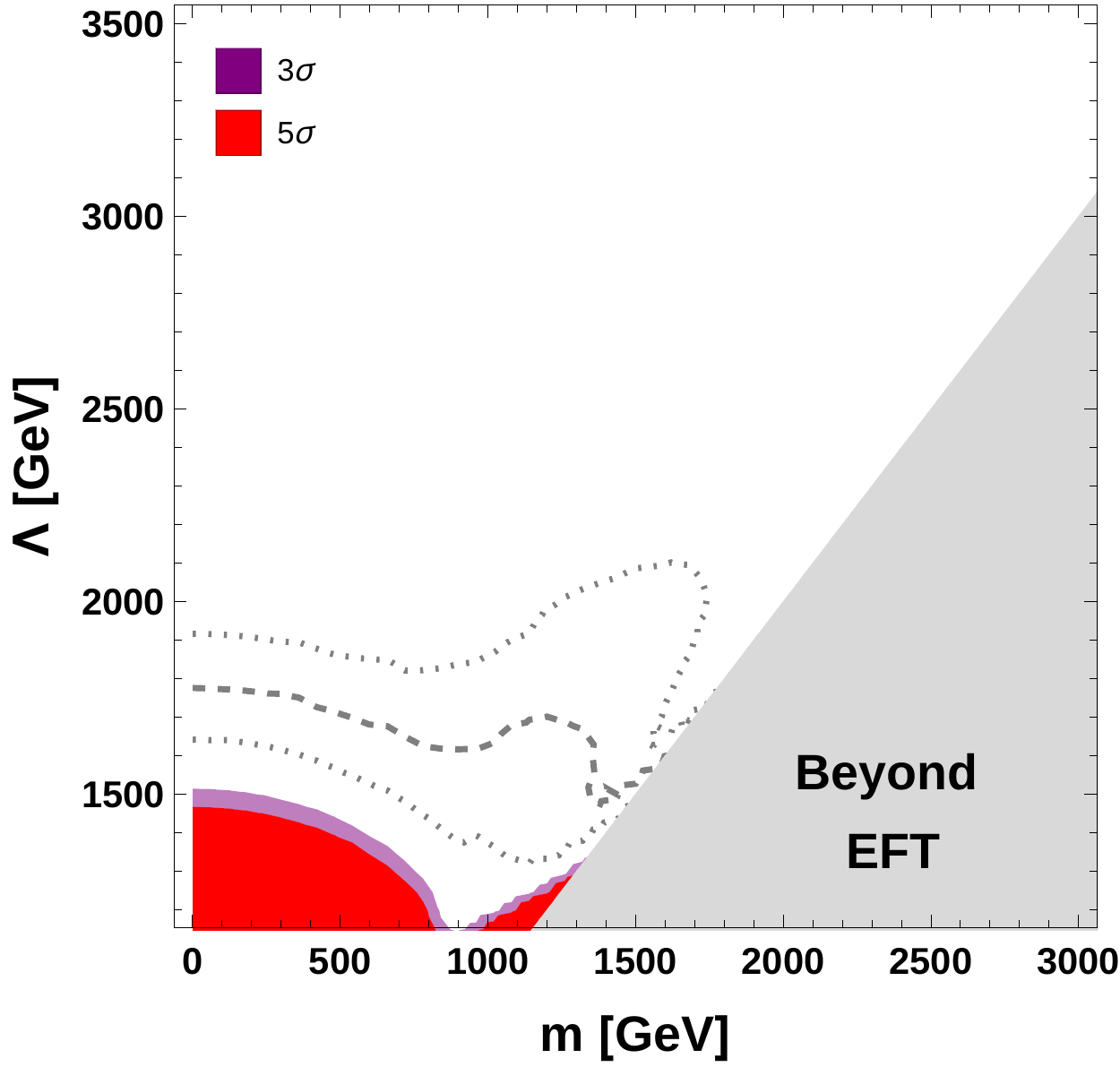}}
\put(230,0){\includegraphics[scale=0.6,clip=true, trim= 0cm 0cm 0cm 0cm]{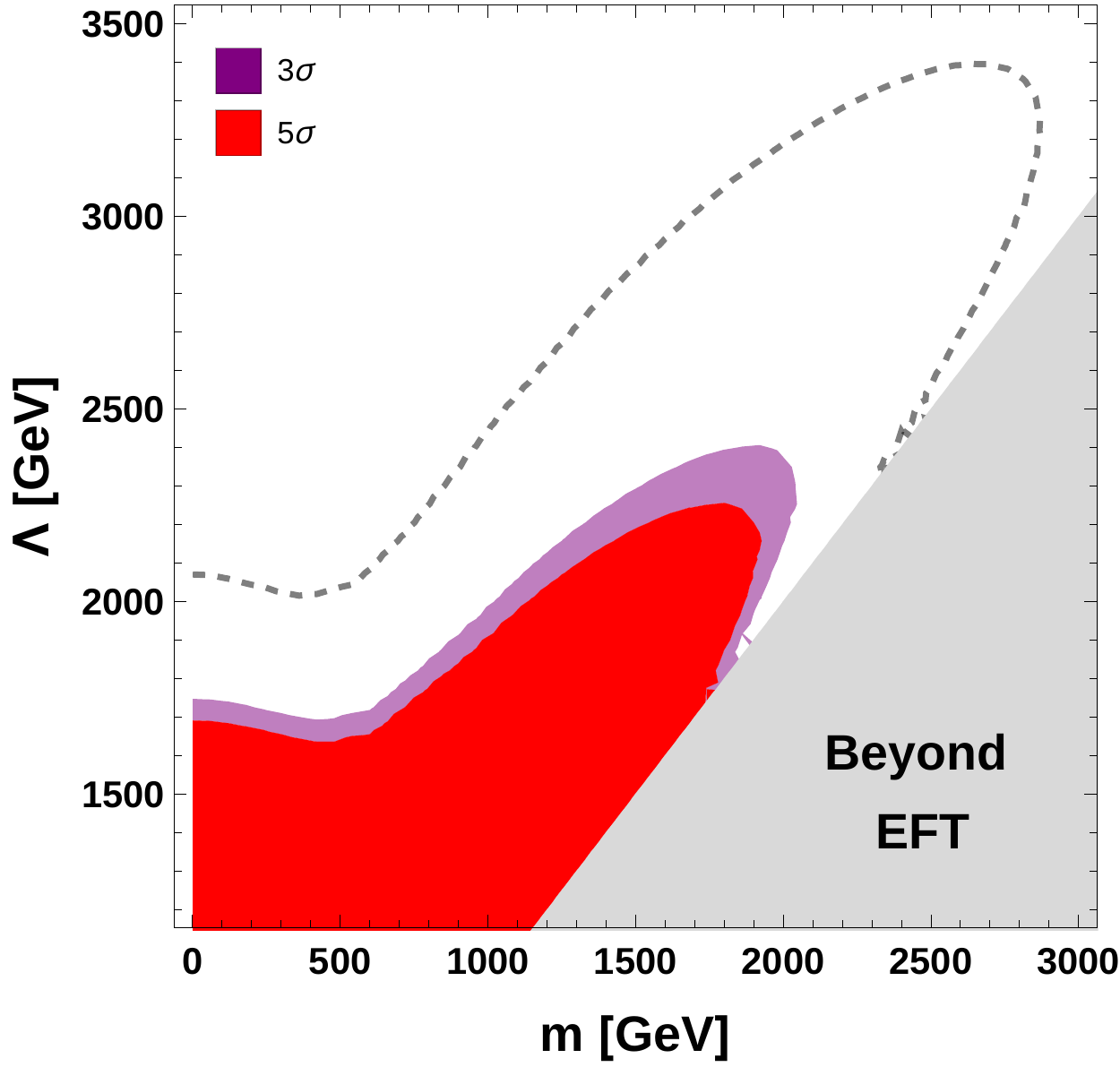}}
\put(80, 185){$\O^0_{8a}$ polarizability}
\put(310, 185){$\O^0_{8b}$ polarizability}
\end{picture}
\caption{Sensitivity of the exclusive diphoton channel to a spin-0 dark particle with dimension-8 polarizability with coefficient $c^0_{8a,8b}(\Lambda)=10$, represented in the mass-cutoff plane, and assuming $\sqrt{s}=13$~TeV, $L=300$~fb$^{-1}$. The dashed lines correspond to the $5~\sigma$ sensitivity in presence of $N=5$ copies of the dark particle. The two dotted lines corresponds to the $5~\sigma$ sensitivities for the spin-0 state of the neutral string assuming $q=1$ and $q=2$, and taking $\Lambda=l_s^{-1}$, .
\label{fig:sens}
}
\end{figure}

The average sensitivities for a signal induced by the $\O^0_{8a}$, $\O^0_{8b}$, $\O^0_{6a}$ operators are shown in Figs.~\ref{fig:sens} and \ref{fig:sens6}. We simply set that   $3~\sigma$ and $5~\sigma$ statistical significance for the existence of a signal roughly correspond  to $\hat n=3$ and $5$ observed events. More evolved statistical analysis  give similar conclusions. 
The uncertainty on the cross-section is expected to blow up when approaching the $m=\Lambda$ limit. In fact, the cross-sections used for  the figures are probably under-estimated in this region because we did not included the local $4\gamma$ operators arising from matching, that should dominate in this region has the $\log(m/\Lambda)$ term becomes small (see also discussion in Sec.~\ref{se:consistency_cal}).

We observe that for the chosen values of $c^0_i$, the sensitivity regions can go above the TeV. 
It turns out  that the sensitivity for the $\O^0_{8b}$ operator is better than for the $\hat\O^0_{6a}$, operator, which itself is better than for the $\O^0_{8a}$ operator. The regions for each operators  have sensibly different shapes. In particular, a sensitivity remains at low mass for the $\O^0_{8a}$, $\O^0_{8b}$ operators, while it vanishes for the  $\hat\O^0_{6a}$ operator. 
These estimations are for a single self-conjugate scalar. An important point to keep in mind is that the search we propose is multiplicity-sensitive: The more the dark sector is populated by polarizable dark particles, the more the sensitivity regions improve. For illustration we show how the regions grow when assuming  $N=5$ particles with same mass and couplings. In this case the photon-photon cross-section is enhanced by a $N^2$ factor.

Although  our present study is limited to the case of one operator turned on at a time, some conclusions can already be drawn regarding some realizations of the spin-0 dark particle. 
In the case of a dark bosonic string computed in Sec.~\ref{se:string}, we only have a $\O^0_{8a}$ polarizability, for which the photon-photon search is the less sensitive.  However, if one identifies $\Lambda=l_s^{-1}$, the coefficient of the operator is  large, $c^0_{8a}=4\pi^2 q^2$ (see Eq.~\eqref{eq:Opstring}). For a charge of $q=1$, the sensitivity reaches $m\sim 2.5$~TeV and $\Lambda\sim 3$~TeV, as shown in Fig.~\ref{fig:sens}.

Regarding the dark spin-0 baryon of the Stealth DM scenario \cite{Appelquist:2015zfa}, only a polarizability of $\O^0_{8a}$ has been considered. However, to the best of our understanding, the $\O^0_{8b}$, $\hat \O^0_{6a}$ operators do not need to be zero, provided that the sum of their coefficients is small (see Sec.~\ref{se:micro}). This may make an important difference in the prospects for the diphoton search, as the sensitivity to $\hat \O^0_{6a}$ and particularly $\O^0_{8b}$ is much better than for the $\O^0_{8a}$ coefficient.
Finally, for a pNGB dark particle, we expect all of the three operators to be non-zero.\footnote{For  the pNGB dark particle, it is not clear to us if the operators should enter in a specific combination.}

The present study, as a proof of principle, is limited to the spin-0 case and to turning on one operator at a time. Given these encouraging first results, it would be worthwhile to go further by computing the loops in presence of all operators at a time. Also, it would be certainly interesting to similarly analyze  the  spin-1/2 and spin-1 cases.

 \subsection{Interplay with other searches for a stable dark particle}

\begin{figure}[t]
\centering
\begin{picture}(500,200)
\put(0,0){\includegraphics[scale=0.6,clip=true, trim= 0cm 0cm 0cm 0cm]{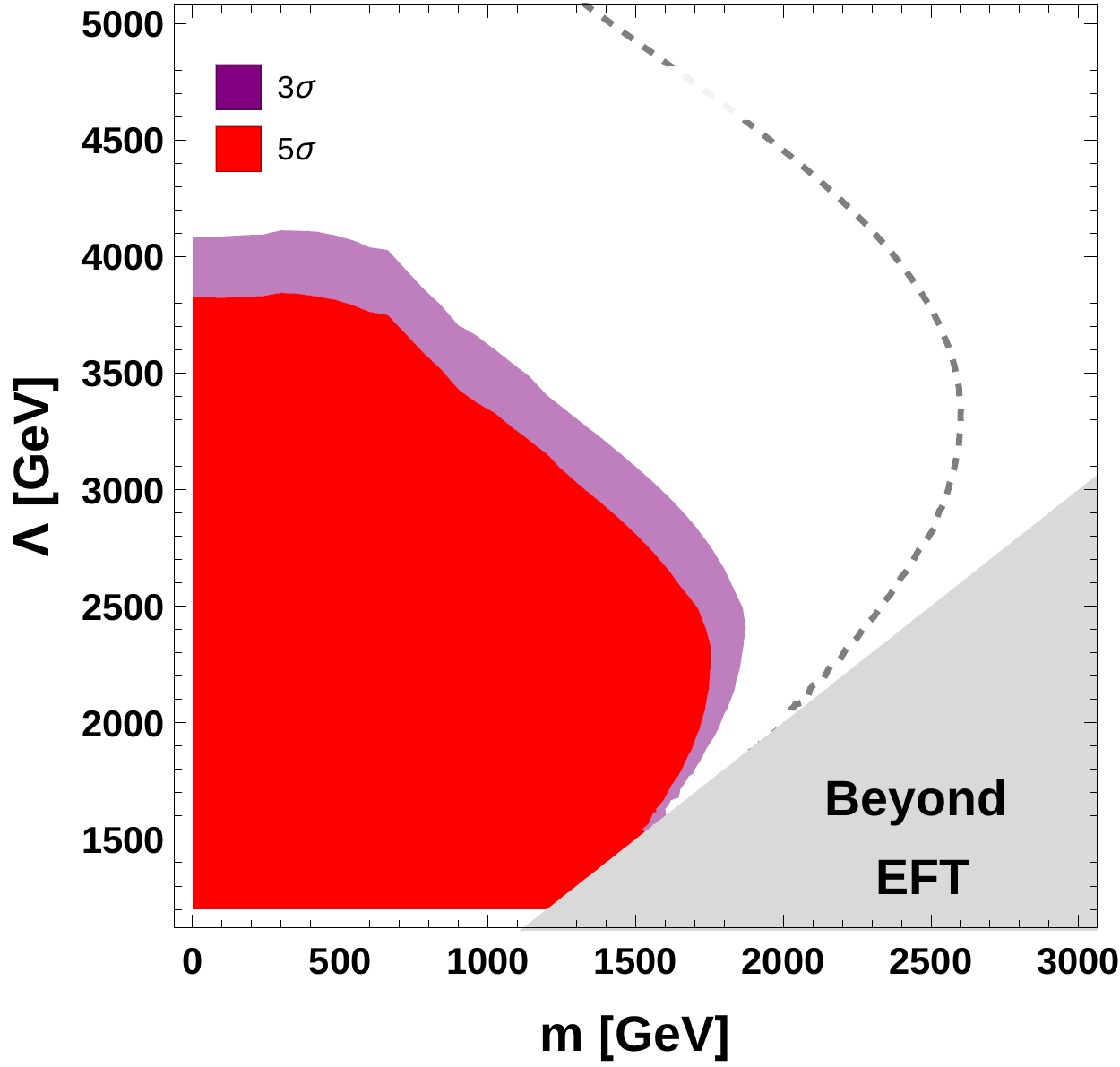}}
\put(230,0){\includegraphics[scale=0.6,clip=true, trim= 0cm 0cm 0cm 0cm]{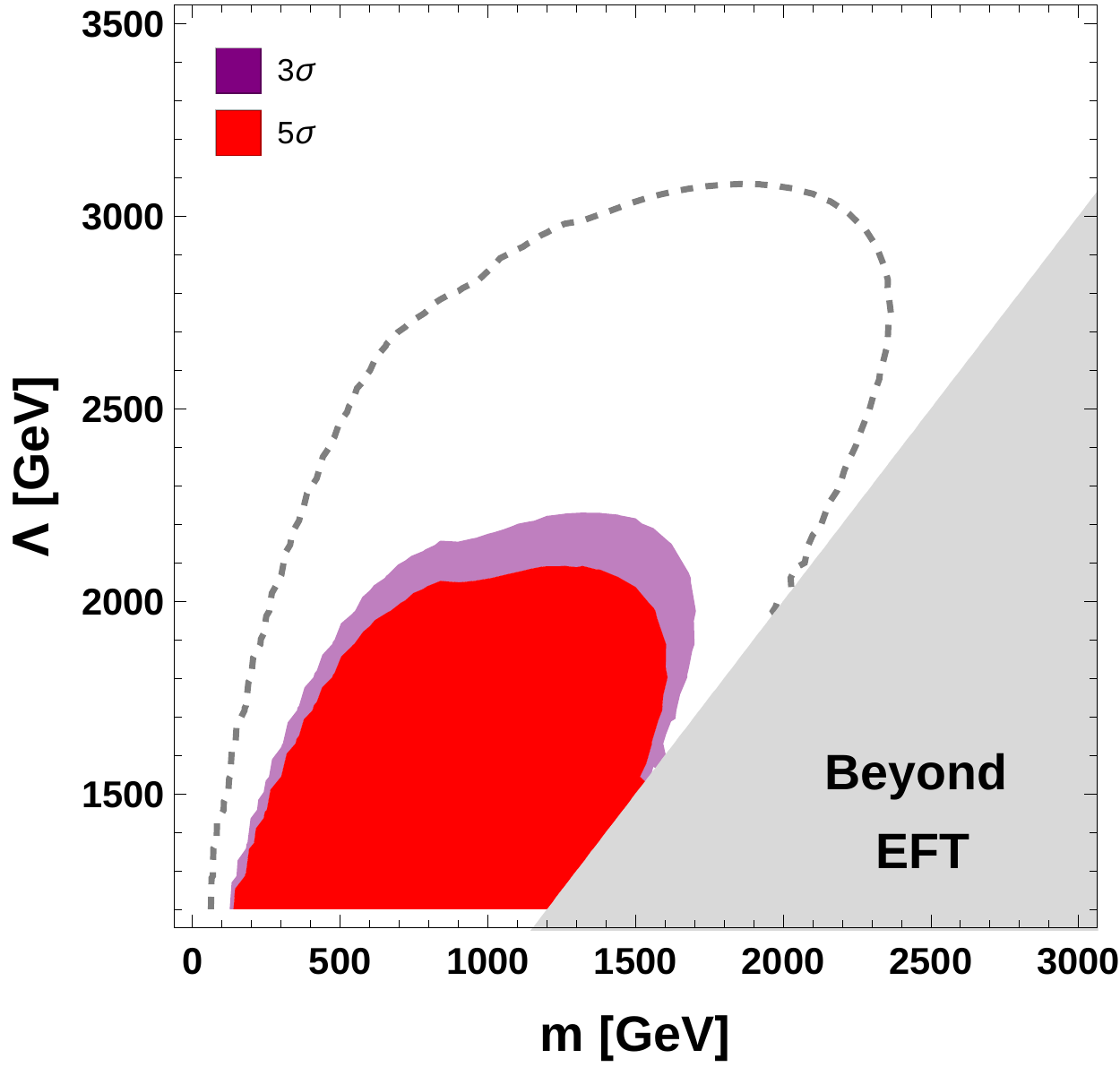}}
\put(80, 185){$ \O^0_{6a}$ polarizability}
\put(310, 185){$\hat \O^0_{6a}$ polarizability}
\end{picture}
\caption{Same as Fig.~\ref{fig:sens}, for dimension-6 polarizability $ \O^0_{6a}$ (left) and dimension-8 polarizability $\hat \O^0_{6a}$ (right) with coefficients $ c^0_{6a}(\Lambda)=10$, $\hat c^0_{6a}(\Lambda)=10$.
\label{fig:sens6}
}
\end{figure}

Here we briefly discuss the case  where the dark particle is stable and identified as  dark matter. We recall that, compared to DM searches, a general drawback of the diphoton search is that it does not detect stability, while a general advantage  is its sensitivity to the entire spectrum of polarizable dark particles.

%

\paragraph{ Comparison with collider searches.}
 A quantitative comparison with the reach of missing-energy searches obtained in the literature (see \textit{e.g.} \cite{Cotta:2012nj, Crivellin:2015wva}) 
would require to take into account the nature of the dark particle, the assumed luminosity and center-of-mass energy, assumptions on the couplings, normalization of the operators and statistical criteria. Here we will remain at a qualitative level.
 We observe that in the prospects for missing-energy based searches at the 13/14 TeV LHC, the sensitivity drops quickly above $m>1$~TeV. While in our case, one can see from Figs.~\ref{fig:sens},~\ref{fig:sens6} that
the sensitivity goes over regions with masses above $\sim 1$~TeV. This can be understood from the kinematics of the two kinds of process: The cross-section for producing two on-shell dark particles plus other states drops faster with the center-of-mass energy than for a photon-photon final state. There is thus a complementarity between the two kind of searches.
 
One can also notice that if the stable dark particle has a multiplicity $N$, the diphoton cross-section grows with $N^2$, but the cross-section for pair production grows only with $N$.
Thus a large multiplicity for the stable dark particle favours the diphoton search, as the photon-photon production is enhanced by $N$ with respect to pair-production. \footnote{ A roughly similar conclusion is expected for $N$ particles which are non-degenerate, as the  decay chains of unstable particles end up  with the stable one and thus contribute to missing energy signatures.}

For these  reasons we conclude that, qualitatively, the proposed diphoton search  seems to compete with and sometimes complement missing-energy  searches at the LHC.

\paragraph{ Comment on indirect detection.}
 A strong constraint on stable polarizable dark particles naturally comes from indirect detection bounds on photons. If the annihilation rate is not velocity-suppressed, these bounds are expected (see \cite{Crivellin:2015wva} and references therein) to dominate over collider and direct searches. 
  As velocity-suppression annihilation is a crucial aspect, we compute the annihilation rate induced simultaneously by the $\O^0_{8a}$ $\O^0_{8b} $ $\hat \O^0_{6a}$ polarizabilities. None of these operators alone lead to a suppressed annihilation rate. However, it turns out that the full squared matrix element takes the form of a complete square
\be
|{\cal M}|^2_{\phi \phi\rightarrow \gamma \gamma }= \frac{32 m^8}{\Lambda^8}\big(c_{8a}-4(c_{8b}+\hat c_{6a}) \big)^2+O\left(\frac{(p_i)^2}{m^2}\right)\,. \label{eq:ann}
\ee
Thus there  exists a combination of coefficients for which the annihilation rate is velocity-suppressed. 
Interestingly, this happens in particular for $\hat c_{6a}=0$, $c_{8b}=c_{8a}/4$, which corresponds precisely to  coupling the traceless part of $F^{\mu\rho}F_\rho^\nu$ to $\partial^\mu \phi \partial^\nu \phi$. \footnote{This is consistent with the velocity-suppressed rate found in \cite{Bruggisser:2016nzw}, Tab.~4. }

 Such operator appears in particular when integrating out a heavy spin-2 particle, like a KK graviton. It would be interesting to further investigate this effective scenario of a ``spin-2 portal''. 
From the point of view of the diphoton search, the  spin-2 particle is a mediator, thus the loop of the polarizable scalar is subdominant with respect to the spin-2 induced four-photon loop.  It would be interesting to investigate whether the combination of Eq.~\eqref{eq:ann} can vanish in a scenario with intrinsic polarizability.

\section{Conclusions } 
\label{se:conclusions}

We propose to test the existence of a  self-conjugate polarizable particle by searching  for the virtual effects it induces. We focus on the process of photon-photon scattering, occuring via  loops of this ``almost dark'' particle.  The method does not depend on whether the  particle is stable.
Thus if there is a dark sector with many polarizable dark particles, the search is sensitive to the cumulative effect of the whole spectrum.

As a preliminary step we classified the CP-even polarizability  operators up to dimension 8 for particles with spin $0$, $1/2$, $1$. We further identified two possible  scenarios  for the microscopic nature of polarizability: mediated and intrinsic polarizability. We illustrate intrinsic polarizability in the case of a neutral bosonic open string and find it is described by dimension-8 operators.  


The scenario of a dark particle with intrinsic polarizability  is the relevant one for the search we propose.  Focussing on the spin-0 case, we evaluate the four-photon helicity amplitudes induced  by the  dimension-8 polarizability operators. The  matching of this effective interaction onto local four-photon  operators for $s\ll m^2$ is also provided.

We then evaluate  the prospects of a $pp\rightarrow \gamma\gamma\, p p $ search at the 13 TeV LHC using forward detectors to characterize the intact protons.  This channel is known for being sensitive to new physics searches.
For operator coefficients equal to $10$, it turns out that the sensitivity in mass and cutoff can go beyond the TeV. For the string with unit charge, mass and inverse string length can be probed up to roughly $1.5$~TeV.    
The center-of-mass energy of the process is typically of $\sim 1$~TeV, hence the EFT expansion is  roughly valid unless the coefficients of the operators get too small.

In case the dark particle is stable, it is a DM candidate. In this context we qualitatively compare  DM collider searches with our  diphoton search.  It turns out that these two methods are fairly complementary, as the diphoton search tends to have a sensitivity to higher masses and is multiplicity-enhanced. The annihilation rate of two dark particles into photons is found to be suppressed if the $c^0_{8a}-4(c^0_{8b}+\hat c^0_{6a})$ combination vanishes. This happens in case of mediated polarizability from a spin-2 particle, and it would be interesting to find a UV completion of intrinsic polarizability in which this cancellation occurs. 

We emphasize that the present study of the spin-0 case should be taken as a proof of concept, used  to get a rough idea of the sensitivities that can be reached. As the  first conclusions seem encouraging, it would be  interesting to further analyze the spin-$0$ case, and to investigate the cases of polarizable self-conjugate particles of spin-$1/2$ and $1$.

\acknowledgments I would like to thank E. Pont\'on,  G. von Gersdorff, R. Mattheus and A. Ballon for useful discussions and
 M. Saimpert  for clarifications on FPMC.  This work was supported by the S\~ao Paulo Research
Foundation (FAPESP) under grants \#2011/11973 and \#2014/21477-2.

 \appendix
 
 \section{Neutral open string  in an electromagnetic background}
\label{app:string}

In this appendix one sets $l_s^2=\frac{1}{2}$.
The string equations of motions in the background field following from Eq.~\eqref{eq:S_string}
are given by 
\be
\ddot X_\mu -X_\mu''=0\,, 
\ee
\be
 X_\mu'=q F_{\mu\nu} \dot X^\nu~~{\rm if~~}\sigma\in\{0,\pi\}\,,
\ee
with $q_0=-q_1=q$.
The background field being antisymmetric, it can be brought into a $2\times2$ block diagonal form by orthogonal transformations, and it is thus enough to focus on two dimensions, taken to be space dimensions with  $\mu=1,2$. One has 
\be
F_{\mu\nu}=\begin{pmatrix}
0 & f \\ -f & 0
\end{pmatrix}\,,\quad {\rm with}~\mu=1,2\,.
\label{eq:string_bg}
\ee
 Is is further convenient to rotate space coordinates  as 
\be X_+=\frac{1}{\sqrt{2}}(X_1+iX_2)\,,\quad X_-=\frac{1}{\sqrt{2}}(X_1-iX_2)\,.\ee
The boundary conditions become simply
\be
 X_+'=-iq f \dot X_+~~{\rm if~~}\sigma\in\{0,\pi\}\,,
\ee
The  oscillator modes are
\be
\psi_n(\sigma,\tau)=\frac{1}{\sqrt{|n|}} \cos\left( n\, \sigma+\gamma\right)e^{-i\,n\tau}\,,
\ee
 where $\gamma=\tan^{-1} (qf) $. The oscillators and the zero mode shown in Eq.~\eqref{eq:string_sol} are orthogonal according to the inner product
 \be
\left \langle \psi_m |\psi_n \right\rangle= \int_0^\pi \frac{d\sigma}{\pi} \, \psi_m^\dagger \Big( \overrightarrow\partial_\tau  - \overleftarrow\partial_\tau +q\,f\,\big( \delta(\sigma-\pi)-\delta(\sigma)\big)\Big)\psi_n=
i\delta_{mn}\, {\rm sgn }(n)\,.
\label{eq:string_inner}
 \ee
 
 One then  introduces the canonical momentum $P_-=\partial {\cal L} / \partial \dot X_+$, giving
\be 
P_-=\frac{1 }{\pi}\dot X_+ + q A_+ (\delta(\sigma)-\delta(\sigma-\pi))
  \ee
To go further, one  uses the approximation that the background field is constant. The potential is then linear in $X_{1,2}$, and one can make the following gauge choice as in \cite{Abouelsaood:1986gd}, 
\be
A_\mu=\frac{1}{2}f \begin{pmatrix}
-X_2 \\ X_1
\end{pmatrix} \,,
\ee
which reproduces well the background field Eq.~\eqref{eq:string_bg} when using the definition $F_{\mu\nu}=\partial_\mu A_\nu-\partial_\nu A_\mu $.
This provides the canonical momentum
\be 
P_-=\frac{1 }{\pi}\dot X_+ -\frac{i}{2} q X_+ \Big(\delta(\sigma)-\delta(\sigma-\pi)\Big)\,.
\label{eq:string_P-}
  \ee

We have then everything to express the operators $x_\pm$, $p_\pm$, $a^{(\dagger)}_n$, in terms of 
$X_\pm$ and $P_\pm$, using the inner product and Eq.~\eqref{eq:string_P-}.  
Using the canonical equal-time commutators for $X_{\pm}$ and $P_\pm$ 
\be\begin{split}
[X_u(\tau,\sigma),X_v(\tau,\sigma')]=0\,,\quad [P_u(\tau,\sigma),P_v(\tau,\sigma')]=0\,,
\\
 [X_u(\tau,\sigma),P_v(\tau,\sigma')]=i \delta_{uv}\delta(\sigma-\sigma')\,,
\end{split}
\ee
we can check that all the operators satisfy well canonical commutation relations. 
Finally, the $L_0$ operator of the Virasao algebra is given by 
\be L_0=\frac{1}{2}\sum_{\mu=1}^2 (\dot X_\mu+ X'_\mu)^2=(\dot X_+ + X'_+)(\dot X_-+ X'_-)\,,
\ee
which gives Eq.~\eqref{eq:L0} using
\be
\dot X_+ + X'_+=e^{-i\gamma}\left[ \frac{p_+}{\sqrt{1+q^2f^2}}
+ \sum_{n=1}^{\infty}\left(a_n e^{-i n (\tau +  \sigma)} - a^\dagger_n e^{i n (\tau +  \sigma)} \right)  
\right]\,,
\ee
after rotating back to $X_{1,2}$ coordinates and putting together all block matrices to restore all dimensions of spacetime.

 \section{Four-photon amplitude calculations}
 \label{app:LbL}
 
 We define  $\Delta=m^2-x(1-x)q^2$. After loop integration,    all the $x$-dependence of the numerators appears via powers of $x(1-x)$ after combination of all terms.
Thus we  introduce a basis of loop functions
\be
f_{n}(q^2,m,\Lambda)=\int_0^1 dx (x(1-x))^n \log\left(\frac{\Delta(q^2)}{\Lambda^2}\right)\,,
\ee
over which all amplitudes decompose. 
One further introduces the combinations 
\be
A(q^2,m,\Lambda)=
( m^4 f_0 -2 m^2 q^2 f_1 +q^4 f_2 )  \,,
\ee
\be
X(q^2,m,\Lambda)=
(3 m^4 + 2 m^2 q^2) f_0 - 
   (30 m^2 q^2 + 2 q^4) f_1 + 
  28 q^4 f_2\,,
\ee
\be
C(q^2,m,\Lambda)=
( 12 m^4+2q^2m^2) f_0 -(32 m^2 q^2+2q^4) f_1 +24q^4 f_2   \,.
\ee
The helicity amplitudes are then given by
\begin{itemize}
\item $\O^0_{8a}$ operator
\be
{\cal M}_{++++}= -\frac{(c^0_{8a})^2}{32 \pi^2\,\Lambda^8} s^2 (X(s,m,\Lambda)+ A(t,m,\Lambda)+ A(u,m,\Lambda) )\,,
\ee
\be
{\cal M}_{++--}= -\frac{(c^0_{8a})^2}{32 \pi^2\,\Lambda^8} ( s^2 \, X(s,m,\Lambda)+t^2 \, X(t,m,\Lambda)+u^2 \, X(u,m,\Lambda) )\,,
\ee
\be
{\cal M}_{+++-}=0\,.
\ee

\item $\O^0_{8b }$ operator
\be
{\cal M}_{++++}= -\frac{(c^0_{8b})^2}{8 \pi^2\,\Lambda^8} s^2 C(s,m,\Lambda)\,,
\ee
\be
{\cal M}_{++--}= -\frac{(c^0_{8b})^2}{8 \pi^2\,\Lambda^8} ( s^2 C(s,m,\Lambda)+t^2 C(t,m,\Lambda)+u^2 C(u,m,\Lambda) )\,,
\ee
\be
{\cal M}_{+++-}=0\,.
\ee

\item $\O^0_{6a }$ operator
\be
{\cal M}_{++++}=-\frac{(c^0_{6a})^2\,s^2}{2\pi^2\,\Lambda^4}  f_0(s,m,\Lambda)\,,
\ee
\be
{\cal M}_{++--}=-\frac{(c^0_{6a})^2}{2\pi^2\,\Lambda^4} ( s^2 f_0(s,m,\Lambda)+ t^2 f_0(t,m,\Lambda)+ u^2 f_0(u,m,\Lambda) )\,,
\ee
\be
{\cal M}_{+++-}=0\,.
\ee
The unpolarized $\gamma\gamma\rightarrow \gamma\gamma$ cross-section is given by 
\be
\frac{d \sigma}{dt}=\frac{1}{16\pi s^2}\Big(
|{\cal M}_{++++}|^2
+|{\cal M}_{++--}|^2
+|{\cal M}_{+-+-}|^2
+|{\cal M}_{+--+}|^2
+4|{\cal M}_{+++-}|^2
\Big)\,.
\ee

\end{itemize}

\bibliographystyle{JHEP} 

\bibliography{Pol_DM}

\providecommand{\href}[2]{#2}\begingroup\raggedright\begin{thebibliography}{10}

\bibitem{Appelquist:2014jch}
{\bf Lattice Strong Dynamics (LSD)} Collaboration, T.~Appelquist et~al., {\it
  {Composite bosonic baryon dark matter on the lattice: SU(4) baryon spectrum
  and the effective Higgs interaction}},  {\em Phys. Rev.} {\bf D89} (2014),
  no.~9 094508, [\href{http://arxiv.org/abs/1402.6656}{{\tt arXiv:1402.6656}}].

\bibitem{Kilic:2009mi}
C.~Kilic, T.~Okui, and R.~Sundrum, {\it {Vectorlike Confinement at the LHC}},
  {\em JHEP} {\bf 02} (2010) 018, [\href{http://arxiv.org/abs/0906.0577}{{\tt
  arXiv:0906.0577}}].

\bibitem{Bagnasco:1993st}
J.~Bagnasco, M.~Dine, and S.~D. Thomas, {\it {Detecting technibaryon dark
  matter}},  {\em Phys. Lett.} {\bf B320} (1994) 99--104,
  [\href{http://arxiv.org/abs/hep-ph/9310290}{{\tt hep-ph/9310290}}].

\bibitem{Pospelov:2000bq}
M.~Pospelov and T.~ter Veldhuis, {\it {Direct and indirect limits on the
  electromagnetic form-factors of WIMPs}},  {\em Phys. Lett.} {\bf B480} (2000)
  181--186, [\href{http://arxiv.org/abs/hep-ph/0003010}{{\tt hep-ph/0003010}}].

\bibitem{Sigurdson:2004zp}
K.~Sigurdson, M.~Doran, A.~Kurylov, R.~R. Caldwell, and M.~Kamionkowski, {\it
  {Dark-matter electric and magnetic dipole moments}},  {\em Phys. Rev.} {\bf
  D70} (2004) 083501, [\href{http://arxiv.org/abs/astro-ph/0406355}{{\tt
  astro-ph/0406355}}]. [Erratum: Phys. Rev.D73,089903(2006)].

\bibitem{Barger:2010gv}
V.~Barger, W.-Y. Keung, and D.~Marfatia, {\it {Electromagnetic properties of
  dark matter: Dipole moments and charge form factor}},  {\em Phys. Lett.} {\bf
  B696} (2011) 74--78, [\href{http://arxiv.org/abs/1007.4345}{{\tt
  arXiv:1007.4345}}].

\bibitem{Banks:2010eh}
T.~Banks, J.-F. Fortin, and S.~Thomas, {\it {Direct Detection of Dark Matter
  Electromagnetic Dipole Moments}},  \href{http://arxiv.org/abs/1007.5515}{{\tt
  arXiv:1007.5515}}.

\bibitem{Cho:2010br}
W.~S. Cho, J.-H. Huh, I.-W. Kim, J.~E. Kim, and B.~Kyae, {\it {Constraining
  WIMP magnetic moment from CDMS II experiment}},  {\em Phys. Lett.} {\bf B687}
  (2010) 6--10, [\href{http://arxiv.org/abs/1001.0579}{{\tt arXiv:1001.0579}}].
  [Erratum: Phys. Lett.B694,496(2011)].

\bibitem{An:2010kc}
H.~An, S.-L. Chen, R.~N. Mohapatra, S.~Nussinov, and Y.~Zhang, {\it {Energy
  Dependence of Direct Detection Cross Section for Asymmetric Mirror Dark
  Matter}},  {\em Phys. Rev.} {\bf D82} (2010) 023533,
  [\href{http://arxiv.org/abs/1004.3296}{{\tt arXiv:1004.3296}}].

\bibitem{Chang:2010en}
S.~Chang, N.~Weiner, and I.~Yavin, {\it {Magnetic Inelastic Dark Matter}},
  {\em Phys. Rev.} {\bf D82} (2010) 125011,
  [\href{http://arxiv.org/abs/1007.4200}{{\tt arXiv:1007.4200}}].

\bibitem{McDermott:2010pa}
S.~D. McDermott, H.-B. Yu, and K.~M. Zurek, {\it {Turning off the Lights: How
  Dark is Dark Matter?}},  {\em Phys. Rev.} {\bf D83} (2011) 063509,
  [\href{http://arxiv.org/abs/1011.2907}{{\tt arXiv:1011.2907}}].

\bibitem{DelNobile:2012tx}
E.~Del~Nobile, C.~Kouvaris, P.~Panci, F.~Sannino, and J.~Virkajarvi, {\it
  {Light Magnetic Dark Matter in Direct Detection Searches}},  {\em JCAP} {\bf
  1208} (2012) 010, [\href{http://arxiv.org/abs/1203.6652}{{\tt
  arXiv:1203.6652}}].

\bibitem{Vecchi:2013iza}
L.~Vecchi, {\it {WIMPs and Un-Naturalness}},
  \href{http://arxiv.org/abs/1312.5695}{{\tt arXiv:1312.5695}}.

\bibitem{Rajaraman:2012db}
A.~Rajaraman, T.~M.~P. Tait, and D.~Whiteson, {\it {Two Lines or Not Two Lines?
  That is the Question of Gamma Ray Spectra}},  {\em JCAP} {\bf 1209} (2012)
  003, [\href{http://arxiv.org/abs/1205.4723}{{\tt arXiv:1205.4723}}].

\bibitem{Rajaraman:2012fu}
A.~Rajaraman, T.~M.~P. Tait, and A.~M. Wijangco, {\it {Effective Theories of
  Gamma-ray Lines from Dark Matter Annihilation}},  {\em Phys. Dark Univ.} {\bf
  2} (2013) 17--21, [\href{http://arxiv.org/abs/1211.7061}{{\tt
  arXiv:1211.7061}}].

\bibitem{Frandsen:2012db}
M.~T. Frandsen, U.~Haisch, F.~Kahlhoefer, P.~Mertsch, and K.~Schmidt-Hoberg,
  {\it {Loop-induced dark matter direct detection signals from gamma-ray
  lines}},  {\em JCAP} {\bf 1210} (2012) 033,
  [\href{http://arxiv.org/abs/1207.3971}{{\tt arXiv:1207.3971}}].

\bibitem{D'Eramo:2014aba}
F.~D'Eramo and M.~Procura, {\it {Connecting Dark Matter UV Complete Models to
  Direct Detection Rates via Effective Field Theory}},  {\em JHEP} {\bf 04}
  (2015) 054, [\href{http://arxiv.org/abs/1411.3342}{{\tt arXiv:1411.3342}}].

\bibitem{Crivellin:2014qxa}
A.~Crivellin, F.~D'Eramo, and M.~Procura, {\it {New Constraints on Dark Matter
  Effective Theories from Standard Model Loops}},  {\em Phys. Rev. Lett.} {\bf
  112} (2014) 191304, [\href{http://arxiv.org/abs/1402.1173}{{\tt
  arXiv:1402.1173}}].

\bibitem{Fedderke:2013pbc}
M.~A. Fedderke, E.~W. Kolb, T.~Lin, and L.-T. Wang, {\it {Gamma-ray constraints
  on dark-matter annihilation to electroweak gauge and Higgs bosons}},  {\em
  JCAP} {\bf 1401} (2014) 001, [\href{http://arxiv.org/abs/1310.6047}{{\tt
  arXiv:1310.6047}}].

\bibitem{Krall:2014dba}
R.~Krall, M.~Reece, and T.~Roxlo, {\it {Effective field theory and keV lines
  from dark matter}},  {\em JCAP} {\bf 1409} (2014) 007,
  [\href{http://arxiv.org/abs/1403.1240}{{\tt arXiv:1403.1240}}].

\bibitem{Weiner:2012cb}
N.~Weiner and I.~Yavin, {\it {How Dark Are Majorana WIMPs? Signals from MiDM
  and Rayleigh Dark Matter}},  {\em Phys. Rev.} {\bf D86} (2012) 075021,
  [\href{http://arxiv.org/abs/1206.2910}{{\tt arXiv:1206.2910}}].

\bibitem{Cotta:2012nj}
R.~C. Cotta, J.~L. Hewett, M.~P. Le, and T.~G. Rizzo, {\it {Bounds on Dark
  Matter Interactions with Electroweak Gauge Bosons}},  {\em Phys. Rev.} {\bf
  D88} (2013) 116009, [\href{http://arxiv.org/abs/1210.0525}{{\tt
  arXiv:1210.0525}}].

\bibitem{Carpenter:2012rg}
L.~M. Carpenter, A.~Nelson, C.~Shimmin, T.~M.~P. Tait, and D.~Whiteson, {\it
  {Collider searches for dark matter in events with a Z boson and missing
  energy}},  {\em Phys. Rev.} {\bf D87} (2013), no.~7 074005,
  [\href{http://arxiv.org/abs/1212.3352}{{\tt arXiv:1212.3352}}].

\bibitem{Crivellin:2014gpa}
A.~Crivellin and U.~Haisch, {\it {Dark matter direct detection constraints from
  gauge bosons loops}},  {\em Phys. Rev.} {\bf D90} (2014) 115011,
  [\href{http://arxiv.org/abs/1408.5046}{{\tt arXiv:1408.5046}}].

\bibitem{Ovanesyan:2014fha}
G.~Ovanesyan and L.~Vecchi, {\it {Direct detection of dark matter
  polarizability}},  {\em JHEP} {\bf 07} (2015) 128,
  [\href{http://arxiv.org/abs/1410.0601}{{\tt arXiv:1410.0601}}].

\bibitem{Crivellin:2015wva}
A.~Crivellin, U.~Haisch, and A.~Hibbs, {\it {LHC constraints on gauge boson
  couplings to dark matter}},  {\em Phys. Rev.} {\bf D91} (2015) 074028,
  [\href{http://arxiv.org/abs/1501.0090}{{\tt arXiv:1501.0090}}].

\bibitem{Appelquist:2015zfa}
T.~Appelquist et~al., {\it {Detecting Stealth Dark Matter Directly through
  Electromagnetic Polarizability}},  {\em Phys. Rev. Lett.} {\bf 115} (2015),
  no.~17 171803, [\href{http://arxiv.org/abs/1503.0420}{{\tt
  arXiv:1503.0420}}].

\bibitem{Brooke:2016vlw}
J.~Brooke, M.~R. Buckley, P.~Dunne, B.~Penning, J.~Tamanas, and M.~Zgubic, {\it
  {Vector Boson Fusion Searches for Dark Matter at the LHC}},  {\em Phys. Rev.}
  {\bf D93} (2016), no.~11 113013, [\href{http://arxiv.org/abs/1603.0773}{{\tt
  arXiv:1603.0773}}].

\bibitem{Aad:2014tda}
{\bf ATLAS} Collaboration, G.~Aad et~al., {\it {Search for new phenomena in
  events with a photon and missing transverse momentum in $pp$ collisions at
  $\sqrt{s}=8$ TeV with the ATLAS detector}},  {\em Phys. Rev.} {\bf D91}
  (2015), no.~1 012008, [\href{http://arxiv.org/abs/1411.1559}{{\tt
  arXiv:1411.1559}}]. [Erratum: Phys. Rev.D92,no.5,059903(2015)].

\bibitem{Aad:2014vka}
{\bf ATLAS} Collaboration, G.~Aad et~al., {\it {Search for dark matter in
  events with a Z boson and missing transverse momentum in pp collisions at
  $\sqrt{s}$=8 TeV with the ATLAS detector}},  {\em Phys. Rev.} {\bf D90}
  (2014), no.~1 012004, [\href{http://arxiv.org/abs/1404.0051}{{\tt
  arXiv:1404.0051}}].

\bibitem{Chatrchyan:2014tja}
{\bf CMS} Collaboration, S.~Chatrchyan et~al., {\it {Search for invisible
  decays of Higgs bosons in the vector boson fusion and associated ZH
  production modes}},  {\em Eur. Phys. J.} {\bf C74} (2014) 2980,
  [\href{http://arxiv.org/abs/1404.1344}{{\tt arXiv:1404.1344}}].

\bibitem{ATLAS:2014wra}
{\bf ATLAS} Collaboration, G.~Aad et~al., {\it {Search for new particles in
  events with one lepton and missing transverse momentum in $pp$ collisions at
  $\sqrt{s}$ = 8 TeV with the ATLAS detector}},  {\em JHEP} {\bf 09} (2014)
  037, [\href{http://arxiv.org/abs/1407.7494}{{\tt arXiv:1407.7494}}].

\bibitem{Khachatryan:2014rwa}
{\bf CMS} Collaboration, V.~Khachatryan et~al., {\it {Search for new phenomena
  in monophoton final states in proton-proton collisions at $\sqrt s =$ 8
  TeV}},  {\em Phys. Lett.} {\bf B755} (2016) 102--124,
  [\href{http://arxiv.org/abs/1410.8812}{{\tt arXiv:1410.8812}}].

\bibitem{Khachatryan:2014rra}
{\bf CMS} Collaboration, V.~Khachatryan et~al., {\it {Search for dark matter,
  extra dimensions, and unparticles in monojet events in proton–proton
  collisions at $\sqrt{s} = 8$ TeV}},  {\em Eur. Phys. J.} {\bf C75} (2015),
  no.~5 235, [\href{http://arxiv.org/abs/1408.3583}{{\tt arXiv:1408.3583}}].

\bibitem{Khachatryan:2014tva}
{\bf CMS} Collaboration, V.~Khachatryan et~al., {\it {Search for physics beyond
  the standard model in final states with a lepton and missing transverse
  energy in proton-proton collisions at sqrt(s) = 8 TeV}},  {\em Phys. Rev.}
  {\bf D91} (2015), no.~9 092005, [\href{http://arxiv.org/abs/1408.2745}{{\tt
  arXiv:1408.2745}}].

\bibitem{Nelson:2013pqa}
A.~Nelson, L.~M. Carpenter, R.~Cotta, A.~Johnstone, and D.~Whiteson, {\it
  {Confronting the Fermi Line with LHC data: an Effective Theory of Dark Matter
  Interaction with Photons}},  {\em Phys. Rev.} {\bf D89} (2014), no.~5 056011,
  [\href{http://arxiv.org/abs/1307.5064}{{\tt arXiv:1307.5064}}].

\bibitem{Lopez:2014qja}
N.~Lopez, L.~M. Carpenter, R.~Cotta, M.~Frate, N.~Zhou, and D.~Whiteson, {\it
  {Collider Bounds on Indirect Dark Matter Searches: The $WW$ Final State}},
  {\em Phys. Rev.} {\bf D89} (2014), no.~11 115013,
  [\href{http://arxiv.org/abs/1403.6734}{{\tt arXiv:1403.6734}}].

\bibitem{Contino:2016jqw}
R.~Contino, A.~Falkowski, F.~Goertz, C.~Grojean, and F.~Riva, {\it {On the
  Validity of the Effective Field Theory Approach to SM Precision Tests}},
  {\em JHEP} {\bf 07} (2016) 144, [\href{http://arxiv.org/abs/1604.0644}{{\tt
  arXiv:1604.0644}}].

\bibitem{Bruggisser:2016nzw}
S.~Bruggisser, F.~Riva, and A.~Urbano, {\it {The Last Gasp of Dark Matter
  Effective Theory}},  \href{http://arxiv.org/abs/1607.0247}{{\tt
  arXiv:1607.0247}}.

\bibitem{Bruggisser:2016ixa}
S.~Bruggisser, F.~Riva, and A.~Urbano, {\it {Strongly Interacting Light Dark
  Matter}},  \href{http://arxiv.org/abs/1607.0247}{{\tt arXiv:1607.0247}}.

\bibitem{Fichet:2013ola}
S.~Fichet and G.~von Gersdorff, {\it {Anomalous gauge couplings from composite
  Higgs and warped extra dimensions}},  {\em JHEP03(2014)102} (2013)
  [\href{http://arxiv.org/abs/1311.6815}{{\tt arXiv:1311.6815}}].

\bibitem{DeSimone:2010tf}
A.~De~Simone, V.~Sanz, and H.~P. Sato, {\it {Pseudo-Dirac Dark Matter Leaves a
  Trace}},  {\em Phys. Rev. Lett.} {\bf 105} (2010) 121802,
  [\href{http://arxiv.org/abs/1004.1567}{{\tt arXiv:1004.1567}}].

\bibitem{Luke:1992tm}
M.~E. Luke, A.~V. Manohar, and M.~J. Savage, {\it {A QCD Calculation of the
  interaction of quarkonium with nuclei}},  {\em Phys. Lett.} {\bf B288} (1992)
  355--359, [\href{http://arxiv.org/abs/hep-ph/9204219}{{\tt hep-ph/9204219}}].

\bibitem{Agashe:2014kda}
{\bf Particle Data Group} Collaboration, K.~A. Olive et~al., {\it {Review of
  Particle Physics}},  {\em Chin. Phys.} {\bf C38} (2014) 090001.

\bibitem{Ferrara:1992yc}
S.~Ferrara, M.~Porrati, and V.~L. Telegdi, {\it {g = 2 as the natural value of
  the tree level gyromagnetic ratio of elementary particles}},  {\em Phys.
  Rev.} {\bf D46} (1992) 3529--3537.

\bibitem{becker2006string}
K.~Becker, M.~Becker, and J.~Schwarz, {\em String Theory and M-Theory: A Modern
  Introduction}.
\newblock Cambridge University Press, 2006.

\bibitem{Burgess:1986dw}
C.~P. Burgess, {\it {Open String Instability in Background Electric Fields}},
  {\em Nucl. Phys.} {\bf B294} (1987) 427--444.

\bibitem{Abouelsaood:1986gd}
A.~Abouelsaood, C.~G. Callan, Jr., C.~R. Nappi, and S.~A. Yost, {\it {Open
  Strings in Background Gauge Fields}},  {\em Nucl. Phys.} {\bf B280} (1987)
  599--624.

\bibitem{Fichet:2013gsa}
S.~Fichet, G.~von Gersdorff, O.~Kepka, B.~Lenzi, C.~Royon, et~al., {\it
  {Probing new physics in diphoton production with proton tagging at the Large
  Hadron Collider}},  \href{http://arxiv.org/abs/1312.5153}{{\tt
  arXiv:1312.5153}}.

\bibitem{Fichet:2014uka}
S.~Fichet, G.~von Gersdorff, B.~Lenzi, C.~Royon, and M.~Saimpert, {\it
  {Light-by-light scattering with intact protons at the LHC: from Standard
  Model to New Physics}},  {\em JHEP} {\bf 02} (2015) 165,
  [\href{http://arxiv.org/abs/1411.6629}{{\tt arXiv:1411.6629}}].

\bibitem{Manohar:1996cq}
A.~V. Manohar, {\it {Effective field theories}},  {\em Lect. Notes Phys.} {\bf
  479} (1997) 311--362, [\href{http://arxiv.org/abs/hep-ph/9606222}{{\tt
  hep-ph/9606222}}].

\bibitem{peskin1995introduction}
M.~Peskin and D.~Schroeder, {\em {An Introduction to Quantum Field Theory}}.
\newblock Advanced book classics. Addison-Wesley Publishing Company, 1995.

\bibitem{Alam:1997nk}
S.~Alam, S.~Dawson, and R.~Szalapski, {\it {Low-energy constraints on new
  physics revisited}},  {\em Phys. Rev.} {\bf D57} (1998) 1577--1590,
  [\href{http://arxiv.org/abs/hep-ph/9706542}{{\tt hep-ph/9706542}}].

\bibitem{Costantini:1971cj}
V.~Costantini, B.~De~Tollis, and G.~Pistoni, {\it {Nonlinear effects in quantum
  electrodynamics}},  {\em Nuovo Cim.} {\bf A2} (1971) 733--787.

\bibitem{atlas}
{ATLAS collaboration, CERN-LHCC-2011-012}, {\it {Letter of intent, Phase-I
  upgrade}}, .

\bibitem{cms}
{CMS and TOTEM collaboration, CERN-LHCC-2014-021}, {\it {CMS-TOTEM Precision
  Proton Spectrometer}}, .

\bibitem{Fichet:2016prl}
S.~Fichet, G.~von Gersdorff, and C.~Royon, {\it {Measuring the Diphoton
  Coupling of a 750 GeV Resonance}},  {\em Phys. Rev. Lett.} {\bf 116} (2016),
  no.~23 231801, [\href{http://arxiv.org/abs/1601.0171}{{\tt
  arXiv:1601.0171}}].

\bibitem{usww}
E.~Chapon, C.~Royon, and O.~Kepka, {\it {Anomalous quartic W W gamma gamma, Z Z
  gamma gamma, and trilinear WW gamma couplings in two-photon processes at high
  luminosity at the LHC}},  {\em Phys.Rev.} {\bf D81} (2010) 074003,
  [\href{http://arxiv.org/abs/0912.5161}{{\tt arXiv:0912.5161}}].

\bibitem{usw}
O.~Kepka and C.~Royon, {\it {Anomalous $W W \gamma$ coupling in photon-induced
  processes using forward detectors at the LHC}},  {\em Phys.Rev.} {\bf D78}
  (2008) 073005, [\href{http://arxiv.org/abs/0808.0322}{{\tt
  arXiv:0808.0322}}].

\bibitem{Gupta:2011be}
R.~S. Gupta, {\it {Probing Quartic Neutral Gauge Boson Couplings using
  diffractive photon fusion at the LHC}},  {\em Phys.Rev.} {\bf D85} (2012)
  014006, [\href{http://arxiv.org/abs/1111.3354}{{\tt arXiv:1111.3354}}].

\bibitem{Sun:2014qoa}
H.~Sun, {\it {Probe anomalous tqγ couplings through single top photoproduction
  at the LHC}},  {\em Nucl.Phys.} {\bf B886} (2014) 691--711,
  [\href{http://arxiv.org/abs/1402.1817}{{\tt arXiv:1402.1817}}].

\bibitem{Sun:2014qba}
H.~Sun, {\it {Large Extra Dimension effects through Light-by-Light Scattering
  at the CERN LHC}},  {\em Eur.Phys.J.} {\bf C74} (2014) 2977,
  [\href{http://arxiv.org/abs/1406.3897}{{\tt arXiv:1406.3897}}].

\bibitem{Sun:2014ppa}
H.~Sun, {\it {Dark Matter Searches in Jet plus Missing Energy in $\rm \gamma p$
  collision at CERN LHC}},  {\em Phys.Rev.} {\bf D90} (2014) 035018,
  [\href{http://arxiv.org/abs/1407.5356}{{\tt arXiv:1407.5356}}].

\bibitem{Sahin:2014dua}
I.~Sahin, M.~Koksal, S.~Inan, A.~Billur, B.~Sahin, et~al., {\it {Graviton
  production through photon-quark scattering at the LHC}},
  \href{http://arxiv.org/abs/1409.1796}{{\tt arXiv:1409.1796}}.

\bibitem{Inan:2014mua}
S.~Inan, {\it {Dimension-six anomalous $tq\gamma$ couplings in $\gamma \gamma$
  collision at the LHC}},  \href{http://arxiv.org/abs/1410.3609}{{\tt
  arXiv:1410.3609}}.

\bibitem{YellowReport}
{\bf LHC Forward Physics Working Group} Collaboration, e.~Royon, C. et~al.,
  {\it {LHC Forward Physics}}, .

\bibitem{friend:2013yra}
D.~d'Enterria and G.~G. da~Silveira, {\it {Observing light-by-light scattering
  at the Large Hadron Collider}},  {\em Phys.Rev.Lett.} {\bf 111} (2013)
  080405, [\href{http://arxiv.org/abs/1305.7142}{{\tt arXiv:1305.7142}}].

\bibitem{Boonekamp:2011ky}
M.~Boonekamp, A.~Dechambre, V.~Juranek, O.~Kepka, M.~Rangel, C.~Royon, and
  R.~Staszewski, {\it {FPMC: A Generator for forward physics}},
  \href{http://arxiv.org/abs/1102.2531}{{\tt arXiv:1102.2531}}.

\bibitem{Budnev:1974de}
V.~Budnev, I.~Ginzburg, G.~Meledin, and V.~Serbo, {\it {The Two photon particle
  production mechanism. Physical problems. Applications. Equivalent photon
  approximation}},  {\em Phys.Rept.} {\bf 15} (1975) 181--281.

\bibitem{private}
Private communication from C. Royon.

\end{thebibliography}\endgroup

\end{document}